\documentclass[12pt]{article}

\usepackage{graphicx}
\usepackage{amsmath}
\usepackage{multirow}
\usepackage{longtable}
\usepackage{caption}

\usepackage{nameref}

\usepackage{pdfpages}

\usepackage{amsfonts}

\usepackage{xcolor}
\pagecolor{white}

\usepackage{float}

\usepackage{braket}
\usepackage{amssymb}
\usepackage{amsthm}
\usepackage[margin=1in,letterpaper]{geometry}
\usepackage{cite}
\usepackage[final]{hyperref}
\hypersetup{
colorlinks=true,
linkcolor=black,
citecolor=blue,
}

\newtheorem{theorem}{Theorem}[section]
\newtheorem{defn}{Definition}[section]
\newtheorem{proposition}{Proposition}[section]

\title{A Mathematical Model of the Cell Cycle: Exploring the Impact of Zingerone on Cancer Cell Proliferation}
\author
{Roumen Anguelov$^1$, Micaela Goddard$^1$, Yvette Hlophe$^2$,\\ Kganya Letsoalo$^2$, June Serem$^3$\\[6pt]
$^1$Department of Mathematics and Applied Mathematics,\\University of Pretoria\\
roumen.anguelov@up.ac.za, micaelajgoddard@gmail.com\\[6pt]
$^2$Department of Physiology, University of Pretoria\\
yvette.hlophe@up.ac.za, u18067507@tuks.co.za\\[6pt]
$^3$ Department of Anatomy, University of Pretoria\\
june.serem@up.ac.za}
\date{}
\begin{document}

\maketitle
\parskip 5mm

\section*{\textit{Abstract}}

This paper presents a mathematical model that explores the interactions between Cyclin-Dependent Kinase 1 (CDK1) and the Anaphase-Promoting Complex (APC) in cancer cells. Through the analysis of a dynamical system simulating the CDK1-APC network, we investigate the system's behavior and its implications for cancer progression and potential therapeutic interventions.
Our findings highlight the critical role of CDK1-APC interactions in regulating the cell cycle and examine the impact of Zingerone, a compound derived from ginger, on modulating the period of the oscillatory dynamics.
These results provide new insights into the potential of Zingerone to influence cell proliferation and offer avenues for less harmful cancer treatments.

The quantitative analysis is conducted by first theoretically deriving the cell viability as a function of time and Zingerone concentration, and then validating this function by using experimental data.

\newpage

\tableofcontents
\newpage

\section{Introduction}\label{sec: intro}

The cell cycle is a fundamental process through which cells grow and divide \cite{cell_cycle1}. Disruptions in cell cycle regulation can lead to uncontrolled cell growth, a hallmark of cancer \cite{maz51}. Among the key regulators of the cell cycle are Cyclin-Dependent Kinase 1 (CDK1) and the Anaphase-Promoting Complex (APC), which drive cells through the cell cycle.

Cancer, a leading cause of death globally, has attracted substantial research attention, including mathematical approaches that offer quantitative analyses across various cancer types and treatment strategies \cite{prof}. In particular, a foundational model described in  \cite{oscillating_model} describes how CDK1 and APC are expected to interact with each other, resulting in oscillatory dynamics. From this model, one may devise strategies to slow cell duplication, a critical factor in cancer progression.

The primary focus of this project is to investigate how reducing the cyclin synthesis rate using the natural compound Zingerone can inhibit melanoma cell viability \cite{zing_general}. The existence of a limit cycle is demonstrated by analysing the foundational model, which describes the activation and inactivation dynamics of CDK1 and APC. A detailed qualitative analysis is performed to validate the system's domain, and the approximate period of the limit cycle is calculated. Thereafter, simplifying the model allows for explicit mathematical solutions to easily be found and parameter constraints applied to the model. From this simplified model, the period of the limit cycle is again determined, shedding light on cell cycle oscillations under various conditions.

Experimental cell viability data was collected at multiple time points and under different Zingerone concentrations. Although some measurement error is inherent, fitting theoretical functions to experimental data helps manage variability, confirming that Zingerone effectively slows cancer cell progression. This underscores the value of mathematical modeling and natural treatment methods in cancer research.

The remainder of this paper is structured as follows: Section \ref{sec: bioback} provides an overview of the cell cycle and key intracellular interactions. Section \ref{sec: mathback} introduces the mathematical model used to represent intracellular dynamics, formulated as a system of ordinary differential equations and analysed in the context of dynamical systems theory. Section \ref{sec: theory} develops a theoretical time-dependent cell viability function and presents experimental data, illustrating both the inherent variability and underlying trends. Additionally, this section discusses model fitting, interpretation of results, the derivation of IC$_{50}$ values, and shows how inhibitory concentrations over time can be calculated for any cell viability level. Finally, the concluding section offers final insights and poses future study directions.

\section{Biological Background}\label{sec: bioback}
\subsection{The Normal Cell Cycle}\label{subsec: cell}

The cell cycle refers to the series of sequential events that cells undergo as they grow and divide. It is divided into two main portions: Interphase, the longest phase of the cell cycle, during which the cell grows and prepares for division, and the Mitotic phase (M phase), where cell division occurs.

\noindent Interphase consists of three sub-phases:
\begin{itemize}
\item G1 phase (first growth phase): the cell undergoes initial growth and prepares for DNA replication.
\item S phase (synthesis phase): DNA replication occurs, resulting in each cell having two identical sets of DNA.
\item G2 phase (second growth phase): the cell continues to grow, checks for any DNA damage that may have occurred during replication, and prepares for entry into the Mitotic phase.
\end{itemize}

\noindent The Mitotic phase (M phase) involves two main processes:
\begin{itemize}
\item Mitosis: the division of the cell’s nucleus and its contents.
\item Cytokinesis: the final division, resulting in two daughter cells, each with its own nucleus.
\end{itemize}

\begin{figure}[H]
\centering
\includegraphics[scale=0.3]{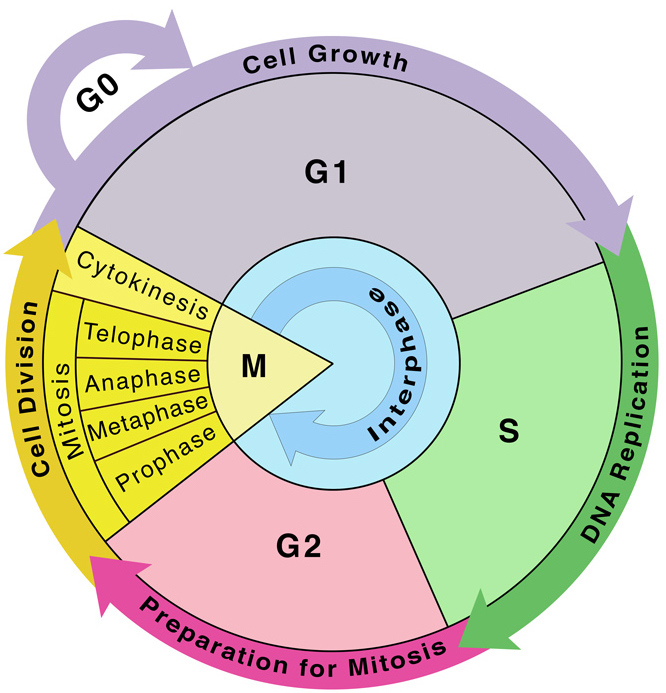}
\caption{Image Source: \cite{cell_cycle_image} -
Progression through the cell cycle. The cell cycle consists of two main phases: Interphase, in which the cell undergoes initial growth (G1), DNA synthesis (S), and secondary growth (G2), and Mitotic phase (M), in which the cell undergoes mitosis and cytokinesis.}
\label{fig: cycle}
\end{figure}

Furthermore, the state G0 in Figure \ref{fig: cycle} represents the process of cells having completed cell division, where the daughter cells enter a new cell cycle in the G1 phase. Alternatively, the cell may withdraw from the cell cycle and be in an out-of-cycle state \cite{cell_cycle1}.
Figure \ref{fig: cycle} illustrates the progression through the cell cycle.

This paper focuses on proliferating cells that actively progress through the cell cycle.
The cell cycle is tightly regulated by sensor mechanisms, called checkpoints, and proteins to ensure that cells only proceed to the next phase when conditions are favourable, and the correct order of events have been maintained. Progression through the cell cycle is driven by Cyclin proteins, which, together with their Cyclin-Dependent Kinase (CDK) enzyme counterpart, tell the cell to continue the cycle. If the checkpoints detect aberrant or incomplete cell cycle events such as DNA damage or cell size abnormalities, checkpoint pathways carry the signal to trigger cell cycle arrest until the problem is resolved \cite{cell_cycle1, cancer1}.

\subsection{Cancer Cells}\label{subsec: cancer}

Disruptions in cell cycle regulation can lead to uncontrolled cell division, a hallmark of cancer. Understanding the cell cycle is crucial for developing treatments for diseases involving abnormal cell proliferation. Cancer is a complex group of diseases characterized by the uncontrolled growth and spread of abnormal cells. If the spread is left uncontrolled, it can be fatal. According to the World Health Organization (WHO) 2020 statistics, cancer was the second leading cause of death globally \cite{maz51}.

Cancer cells differ from normal cells in several important characteristics. Unlike normal cells, which grow and divide in a regulated manner, cancer cells grow and divide uncontrollably due to mutations in genes that control the cell cycle. Cancer cells can invade nearby tissues and spread to other parts of the body via blood and lymphatic systems, known as metastasis. Cancer cells often evade the process of apoptosis, which is programmed cell death that occurs in normal cells when they are damaged or no longer needed, allowing cancer cells to survive longer than normal cells. They also possess the ability to proliferate indefinitely \cite{cancer1}.

Cancer is classified based on the tissue or organ of origin and the type of cells involved. The major categories include:
\begin{itemize}
\item Carcinomas: Cancers that are found in tissues known as Epithelial tissue, which cover surfaces of organs, glands, or body structures.
\item Sarcomas: Malignant tumors that originate in connective tissues, such as cartilage, bone, muscle, and fat.
\item Leukemias: Cancers found in blood and bone marrow, characterized by the overproduction of abnormal white blood cells, which are not able to fight infection and impair the ability of the bone marrow to produce red blood cells and platelets.
\item Lymphomas: Cancers that originate in the lymphatic system, a key part of the immune system.
\end{itemize}
For further information, the reader is referred to \cite{cancer_types}.

Cancer arises due to mutations in cellular DNA, which can result from numerous factors, including genetic predisposition, environmental exposure, lifestyle choices, and infections. Understanding cancer's complex nature is essential for developing better diagnostic tools, treatments, and prevention strategies. Continued research is vital for advancing our knowledge and improving outcomes for those affected by cancer.

\subsection{Natural Treatments as Alternatives to Conventional Cancer Therapies}\label{subsec: treatment}

Cancer treatment typically depends on the type and stage of the disease, with common options including surgery to remove tumors, radiation therapy to destroy or damage cancer cells, or chemotherapy drugs to stop the growth of cancer cells. While these conventional treatments have significantly improved the survival rates for many types of cancer, they come with major drawbacks. Common side effects associated with these treatments include pain, infection, fatigue, and nausea.

Given these side effects, there is growing interest in exploring natural treatments for cancer as potentially less harmful alternatives. These approaches tend to be less toxic and may improve the quality of life for patients. Often, natural treatment approaches have fewer or more manageable side effects, or are used in conjunction with conventional treatments to help manage symptoms and improve outcomes. Additionally, some natural treatments focus on the prevention of cancer and support the body's natural defense against cancer development \cite{zing_med}. Several natural remedies and plants have been studied for their potential anti-cancer properties. While ongoing research into natural remedies continues, some compounds have shown promise in laboratory studies, though they should not be viewed as replacements for standard treatments.

A few notable examples of natural remedies and plants which display anti-cancer properties include:

\begin{itemize}
\item \textbf{Turmeric} (Curcumin): \newline
Multiple reports have been published demonstrating the anticancer potential of curcumin. Studies show that curcumin can inhibit the growth of various types of cancer cells \cite{curcumin}.

\item \textbf{Green Tea} (Catechins, particularly epigallocatechin gallate (EGCG)): \newline
EGCG, a powerful antioxidant, has demonstrated prominent anti-cancer effects \cite{egcg}.

\item \textbf{Ginger} (Gingerol and Zingerone): \newline
Ginger compounds have been shown to induce apoptosis (programmed cell death) and inhibit the growth of cancer cells \cite{zing_general}.
\end{itemize}

This discussion focuses specifically on Zingerone, a compound derived from ginger, and its effects on melanoma cancer cells. Ginger has long been recognized for its medicinal properties and is rich in essential vitamins and minerals. Traditionally used to treat various ailments, ginger contains active ingredients with numerous pharmacological benefits \cite{zing}. Zingerone, in particular, is a non-toxic compound that exhibits anti-cancer properties, along with other effects such as anti-inflammatory, antidiabetic, antidiarrheal, and antispasmodic actions. Notably, Zingerone is derived from natural sources and has no known side effects \cite{maz51}.

\subsection{Interactions Between Cyclin-Dependent Kinase 1 (CDK1) and Anaphase-Promoting Complex (APC)}\label{subsec: APC CDK1}

This review explores the abnormalities in cell cycle regulatory proteins and their potential implications for cancer treatment. To understand how these disruptions contribute to cancer, we must first examine the role of these proteins in the normal cell cycle \cite{cancer1}. The interaction between Cyclin-Dependent Kinase 1 (CDK1) and the Anaphase-Promoting Complex (APC) is essential for cell cycle regulation, particularly during the transition from mitosis to the G1 phase.

\subsubsection{CDK1 and its Role in the Cell Cycle}

\begin{figure}[H]
\centering
\includegraphics[scale=0.6]{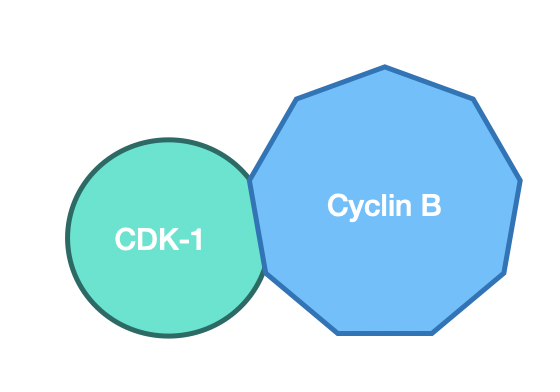}
\caption{A simplified diagram illustrating CDK1 activation through its binding to Cyclin B.}
\label{fig: cdk1}
\end{figure}

CDK1 is a pivotal Cyclin-Dependent Kinase responsible for driving the cell cycle forward. In the G2 phase, CDK1 binds to the Cyclin-B protein to become active, as seen in Figure \ref{fig: cdk1}. This complex, known as the Maturation-Promoting Factor (MPF), triggers CDK1 activation, enabling the cell to enter mitosis \cite{cell_cycle1}.

\subsubsection{APC and its Role in the Cell Cycle}

The APC is crucial for the regulation of the Metaphase-to-Anaphase transition and subsequent exit from Mitosis. The APC targets Cyclin B for degradation, leading to the inactivation of CDK1. This degradation is crucial for exit from mitosis and progression into G1.
Without the degradation of Cyclin B, the cell would be unable to complete mitosis.

\begin{figure}[H]
\centering
\includegraphics[scale=0.6]{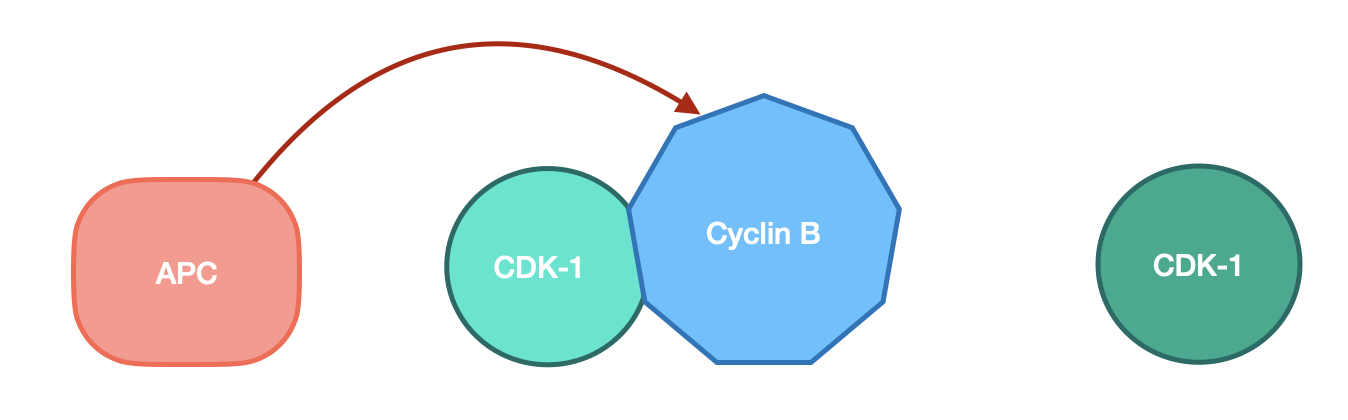}
\caption{A simplified diagram depicting APC targeting Cyclin B for degradation, resulting in CDK1 inactivation.}
\label{fig: apc}
\end{figure}

\subsubsection{Interactions between CDK1 and APC}

The cell cycle is driven by a protein circuit centered on the cyclin-dependent kinase (CDK1) and the anaphase-promoting complex (APC). The carefully coordinated interaction between CDK1 and APC maintains the orderly progression through the cell cycle. The activation of CDK1 drives the cell into mitosis, whereas the activation of APC, which generally lags behind CDK1, facilitates exit from mitosis and entry into G1. Disruption in these processes can lead to cell cycle defects, contributing to the development of cancer and other diseases.

There are still some missing components and poorly understood connections, but overall, the cell-cycle network is fairly well mapped out. Research continues to uncover the finer details of how CDK1 and APC regulate the cell cycle and how their dysfunction may lead to cancer \cite{oscillating_model}. Understanding these interactions, especially the differences between normal and cancerous cells, is critical for developing targeted therapies and improving patient outcomes.

\section{Mathematical Background and Modeling}\label{sec: mathback}

\subsection{Dynamical Systems and Limit Cycles}\label{subsec: theory}

Mathematical modeling plays a crucial role in understanding biological systems, including cancer cell interactions. It allows researchers to simulate complex biological processes, predict system behaviour, and gain insights that may not be easily attainable through experimental methods alone. Mathematical models offer a way to test biological hypotheses in silico, reducing the need for time-consuming and costly experiments.

Mathematical modeling is an invaluable tool in cancer research, providing insights into complex systems and guiding experimental and clinical efforts. As computational power and data availability continue to increase, the role of mathematical models in understanding and treating cancer is expected to expand further.

A dynamical system is a concept used to describe a system that evolves over time. Dynamical systems theory has become a valuable tool to analyse and understand a wide range of problems. The field now includes a well-developed qualitative approach to dynamical systems, usually based on the system's evolution being defined by a smooth function of its arguments. This approach has proved highly effective in helping to understand the behavior of many important physical phenomena, including biological systems \cite{piece_smooth}.

Consider the following initial value problem:

\begin{align} {u}'(t) = \bar{v}({u}(t)) \text{ with } {u}(0) = {U} \in \mathbb{R}^p \label{ivp} \end{align}
where ${u}(t) \in \mathbb{R}^p $ denotes a vector valued function of $t \in \mathbb{R}$, and we assume that $\bar{v} \in C(\mathbb{R}^p, \mathbb{R}^p)$.

\begin{defn}
[\normalfont{Dynamical System}] \label{def: dyn sys} $\newline$
Equation \normalfont{(\ref{ivp})} \textit{defines a dynamical system on a subset $B \subset \mathbb{R}^p$ if, for every $U \in B$, there exists a unique solution of} (\ref{ivp}) \textit{which is defined for all $t \in [0,\infty)$ and remaining in $B$ for all  $t \in [0,\infty)$ }\cite{stuart}.
\end{defn}
\vspace{2mm}
\begin{defn}
[\normalfont{Flow-Invariance}] \label{def: invariant} $\newline$
A set $B \subset \mathbb{R}^p$ is said to be (positively) invariant or flow-invariant with respect to the system $ {u}'(t) = \bar{v}({u}(t)) $ if for any solution $u(t)$, $u(a) \in B$ implies $u(t) \in B$ for $t>a$, as long as the solution exists \cite{walter}.
\end{defn}
\vspace{2mm}
\begin{theorem}
[\normalfont{Tangent Condition}] \label{thm: tangent} $\newline$
The basic hypothesis for invariance is a tangent condition that roughly states that at a boundary point $z \in \partial B$ the vector $\bar{v}(z)$ is either tangent to $B$ or points into the interior of $B$.

The tangent condition is given as follows \normalfont \cite{walter}:
\begin{align} \langle n(z), \bar{v}(z) \rangle \leq 0 \textit { for each } z \in \partial B, \textit{where $n(z)$ is the outer normal to $B$ at $z$}. \label{tangent} \end{align}
\end{theorem}
\vspace{2mm}
\begin{proposition}
[\normalfont{Locally Lipschitz Condition}] \label{prop: ll} $\newline$
A function $\bar{v}:  \mathbb{R}^p \to \mathbb{R}^p$ is locally Lipschitz on an open set $D \subset \mathbb{R}^p$ if the partial derivatives of its components are continuous on $D$ \normalfont \cite{unique}.
\end{proposition}
\vspace{2mm}
\begin{theorem}
[\normalfont{Picard-Lindelöf Theorem}] \label{thm: pl} $\newline$
Consider the autonomous system $ {u}'(t) = \bar{v}({u}(t)) $ and suppose the function $\bar{v}$ is locally Lipschitz on its domain. If $u$ and $w$ are solutions of the system defined on $(\alpha, \beta)$ and $u(s) = w(s)$ for some $s \in (\alpha, \beta)$ then $u=w$ on $(\alpha, \beta)$ \normalfont\cite{unique}.
\end{theorem}
\vspace{2mm}
\begin{theorem}
[\normalfont{Invariance Theorem}] \label{thm: invar} $\newline$
The closed set $B \subset \mathbb{R}^p$ is flow-invariant with respect to the system $ {u}'(t) = \bar{v}({u}(t)) $ if $\bar{v}$ satisfies the tangent condition \normalfont(\ref{tangent})\textit{ and is locally Lipschitz} \cite{walter}.
\end{theorem}
\vspace{2mm}
\begin{theorem}
[\normalfont{Local Existence}] \label{thm: le} $\newline$
Consider the initial value problem in (\ref{ivp}).
Suppose the function $\bar{v}$ is locally Lipschitz on its domain $B$. Then, for any $U \in B$, there exists a positive number $\delta$ and a function $u$ such that $u$ is a solution of the initial value problem on $(-\delta, \delta)$\normalfont \cite{unique}.
\end{theorem}
\vspace{2mm}
\begin{theorem}
[\normalfont{Existence and Uniqueness for Locally Lipschitz Problems}] \label{thm: dyn_sys} $\newline$
Suppose the function $\bar{v}$ is locally Lipschitz on a bounded subset $B$ of its domain $D$, $B \subset D \subset \mathbb{R}^p$. If it may be shown that for any $U \in B$, the solution $u(t)$ of (\ref{ivp}) satisfies $u(t) \in B$ for each $t \geq 0$ such that the solution exists, then (\ref{ivp}) defines a dynamical system on $B$ \normalfont \cite{stuart}.
\end{theorem}
\vspace{2mm}

Limit cycles are an important concept in the study of dynamical systems, particularly for non-linear differential equations. A limit cycle is a closed trajectory in the phase space of a dynamical system such that at least one other trajectory spirals into it as time approaches infinity. A limit cycle is considered stable if all surrounding trajectories spiral inward toward it over time.
$\\$

Let $(x(t), y(t))$ be a solution to a dynamical system that is bounded as $t \rightarrow \infty$.
\vspace{2mm}
\begin{defn}
[\normalfont{Positive Semi-orbit}] $\newline$
The positive semi-orbit $C^+$ of this solution is defined to be the set of points $(x(t), y(t))$ for $t \geq 0$ in the $xy$-plane \cite{772}.
\end{defn}
\vspace{2mm}
\begin{defn}
[\normalfont{Limit Set}] $\newline$
The limit set $L(C^+)$ of the semi-orbit $C^+$ is defined to be the set of all points $(\bar x(t), \bar y(t))$ such that there is a sequence of times $t_n \rightarrow \infty$ with $x(t_n) \rightarrow \bar x$ and $y(t_n) \rightarrow \bar y$ as $n \rightarrow \infty$ \cite{772}.
\end{defn}
\vspace{2mm}
\begin{theorem}
[\normalfont{Poincar\'e Bendixson Theorem}] \label{thm: pb} $\newline$
Assume that the solutions of the system exist and are unique. If $C^+$ is a bounded semi-orbit whose limit set $L(C^+)$ contains no equilibrium points, then either $C^+$ is a periodic orbit and $L(C^+) = C^+$, or $L(C^+)$ is a periodic orbit, called a limit cycle, which $C^+$ approaches spirally, either from inside or outside \cite{772}.
\end{theorem}

In the special case of only one equilibrium point, the Poincar\'e Bendixson Theorem asserts that if the system's forward-time trajectories are bounded and the equilibrium point is unstable, then a limit cycle must exist.

\subsection{Mathematical Model - A Dynamical System Describing the Interactions of CDK1 and APC}\label{subsec: model}

The interactions between CDK1 and APC are discussed, and a two ordinary differential equation model with interlinked positive and negative feedback loops between CDK1 and APC is presented in \cite{oscillating_model}, which enabled an understanding of why certain cell cycles oscillate.

The Hill Function is a sigmoidal function often used in biochemistry, which models cooperative interactions. The Hill function is given by:
\begin{equation}
h(K, x) = \frac{x^{n}}{K^{n} + x^{n}} \nonumber
\end{equation}
where $x$ is the level of activation of the protein, $n$ is the Hill coefficient, and $K$ is a constant. The range of the Hill function is \( 0 \leq h(K, x) \leq 1 \).

The behaviour of the Hill function can be summarised as follows:
\begin{itemize}
\item When \( x \ll K \), \( h(K, x) \approx 0 \), so there is little binding or activation.
\item When \( x \gg K \), \( h(K, x) \approx 1 \), the system becomes saturated, and the function approaches its maximum.
\item When \( n \to \infty \), the Hill function is approximately equal to the Heaviside Function $H(K,x)$ (a step function):
  \[
  h(K, x) \approx H(K,x)=
  \begin{cases}
  0 & \text{if } x < K \\
  1 & \text{if } x > K
  \end{cases}
  \]
 The function becomes almost like an "on-off" switch. This behavior is useful for modeling situations where a system transitions between two states sharply as the level of activation $x$ passes a critical threshold.
\end{itemize}

\subsubsection*{Variables and Notation}\label{subsubsec: vars}

\begin{figure}[H]
\centering
\includegraphics[scale=0.65]{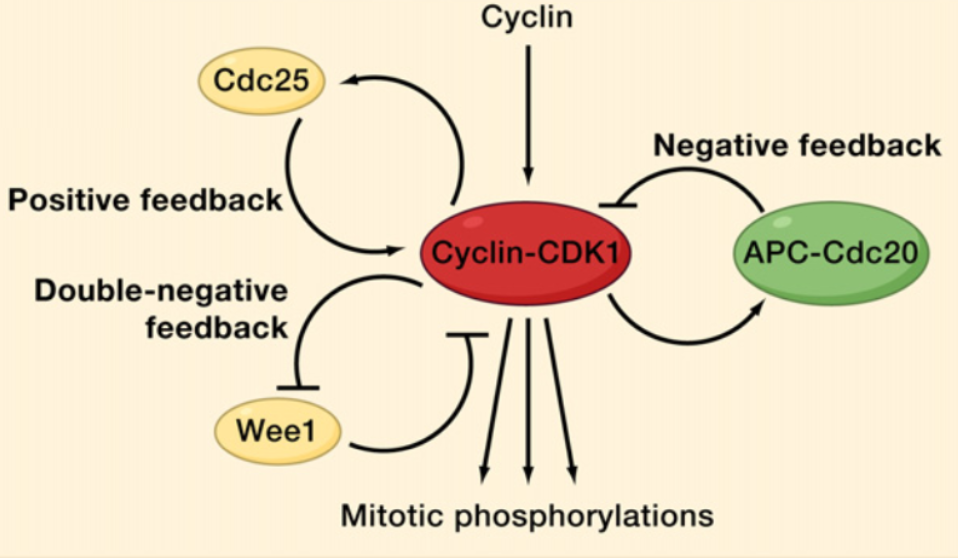}
\caption{Image Source: \cite{oscillating_model} - This Figure illustrates a simplified depiction of the cell cycle. This figure highlights the main interactions between CDK1 and APC. For further details, the reader is referred to \cite{oscillating_model}.}
\label{Fig: interaction}
\end{figure}

Based on the interaction network in Figure \ref{Fig: interaction}, the following variables and notations are used in the model \cite{oscillating_model}:

\begin{itemize}
\item $t$ denotes the time on which the system is solved. The domain of $t$ is the interval from $0$ to infinity (forward-time domain).
\item $[CDK1](t)$ denotes the level of activation of CDK1 at time $t$.
\item $[APC](t)$ denotes the level of activation of APC at time $t$.
\item $\alpha_1$ denotes a rate constant of Cyclin synthesis.
\item $ - \beta_1 [\text{CDK1}]  h(K_1, [\text{APC}])$ represents the deactivation of CDK1 via APC (through Cyclin B1 degradation), the rate of which should be proportional to a Hill function of APC, with a Hill coefficient of $n_1$.
\item $ \alpha_3 (1-[\text{CDK1}]) h(K_3, [\text{CDK1}])$ is an additional positive feedback term for the self-promotion of CDK1, the rate of which should be proportional to a Hill function of CDK1, with a Hill coefficient of $n_3$.
\item $\alpha_2 (1-[\text{APC}])h(K_2, [\text{CDK1}])$ accounts for the rate of the activation of APC under the assumption that the activation rate is proportional to the level of activation (1 – [APC]) of the inactive APC multiplied by the Hill function of CDK1, with a Hill coefficient of $n_2$.
\item $- \beta_2 [\text{APC}] $ represents the inactivation rate of APC, which is proportional to [APC].
\end{itemize}

\subsubsection*{The Full Model}\label{subsec: fullmodel}

The two-ODE model that was proposed follows:

 \begin{align}
 \frac{d[\text{CDK1}]}{dt} &= \alpha_1 - \beta_1 [\text{CDK1}] \frac{[\text{APC}]^{n_1}}{K_1^{n_1} + [\text{APC}]^{n_1}} + \alpha_3 (1-[\text{CDK1}]) \frac{[\text{CDK1}]^{n_3}}{K_3^{n_3} + [\text{CDK1}]^{n_3}} \label{Eq: 1}\\
 \frac{d[\text{APC}]}{dt} &= \alpha_2 (1-[\text{APC}]) \frac{[\text{CDK1}]^{n_2}}{K_2^{n_2} + [\text{CDK1}]^{n_2}} - \beta_2 [\text{APC}] \label{Eq: 2}
 \end{align}

\noindent The only set of points for the system that is physically plausible is the set
\begin{equation}
\Omega := \{ ([\text{CDK1}], [\text{APC}]) : 0 \leq [\text{CDK1}] \leq 1, 0 \leq [\text{APC}] \leq 1 \}, \label{border}
\end{equation}
since [CDK1] and [APC] represent the level of activation of the respective proteins \cite{oscillating_model}. Here, a level of activation equal to $1$ indicates that all proteins in the system are active, while a level of activation equal to $0$ indicates that all proteins in the system are inactive.

\begin{figure}[H]
\centering
\includegraphics[scale=0.65]{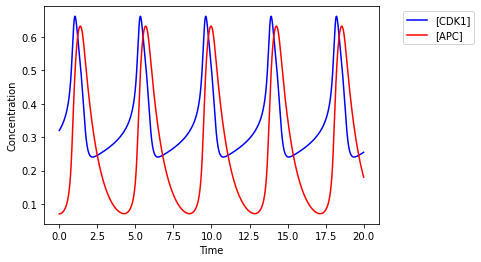}
\caption{Solution curves displaying the level of activations of CDK1 and APC over time.}
\label{fig: original}
\end{figure}

\begin{figure}[H]
\centering
\includegraphics[scale=0.65]{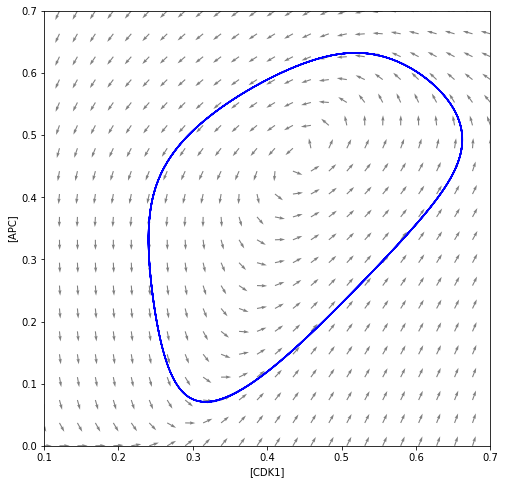}
\caption{Phase portrait and trajectory along a limit cycle of the proposed dynamical system.}
\label{fig: phase}
\end{figure}

Unless otherwise noted, parameters identical to those that were used in previous studies are employed: $\alpha_1 = 0.02, \alpha_2 = 3, \alpha_3 = 3, \beta_1 = 3,\beta_2 = 1, K_1 = 0.5, K_2 = 0.5, K_3 = 0.5, n_1 = 8, n_2 = 8, $ and $n_3 = 8$ \cite{model}. Using these parameters and the initial condition $([CDK1], [APC]) = (0.32, 0.07)$, the solutions for [CDK1] and [APC] are displayed in Figure \ref{fig: original}. Additionally, the phase portrait is sketched in Figure \ref{fig: phase}.

\subsection{Determining the Domain of the Dynamical System} \label{subsec: domain}

For notational simplicity, the level of activation of CDK1 ([CDK1]) will be referred to as $x$, and the level of activation of APC ([APC]) will be referred to as $y$. Likewise, the rate of change of the level of activation of CDK1 $\biggl{(} \frac{d[\text{CDK1}]}{dt} \biggr{)} $ will be referred to as $g$, while the rate of change of the level of activation of APC $\biggl{(} \frac{d[\text{APC}]}{dt} \biggr{)} $ will be referred to as $f$, which are both functions of $x$ and $y$. Equations \ref{Eq: 1} and \ref{Eq: 2} then become:

\begin{align}
\frac{dx}{dt} = g(x,y) &:= \alpha_1 - \beta_1 x \frac{y^{n_1}}{K_1^{n_1} + y^{n_1}} + \alpha_3 (1-x) \frac{x^{n_3}}{K_3^{n_3} + x^{n_3}} \label{Eq: 3}\\
\frac{dy}{dt} = f(x,y) &:=  \alpha_2 (1 - y) \frac{x^{n_2}}{K_2^{n_2} + x^{n_2}} - \beta_2 y \label{Eq: 4}
\end{align}

Notably, $g(x,y)$ and $f(x,y)$ are locally Lipschitz (Appendix \nameref{A5} - the \nameref{prop: ll} is satisfied). To determine whether the system's trajectories remain within the feasible set described in (\ref{border}), an analysis is conducted on the boundary $\partial \Omega$ using the \nameref{thm: tangent} (\ref{tangent}). Let $\bar{v} = \begin{pmatrix} \frac{dx}{dt} \\ \\
\frac{dy}{dt}\end{pmatrix} $. Then,

\begin{itemize}
\item For each point $z$ along the boundary of $\Omega$ where $y=0$; $z\in \partial \Omega \big|_{y=0}$, the outward normal vector is $n(z) = \begin{pmatrix} 0 \\ -1 \end{pmatrix}$. Therefore, \\
\begin{align}
\left\langle n(z), \bar{v}(z) \right\rangle
&= -\alpha_2 \frac{x^{n_2}}{K_2^{n_2} + x^{n_2}} \leq 0 \qquad \forall x \in [0,1] \label{y0_border} \end{align}
\item For each point $z$ along the boundary of $\Omega$ where $y=1$; $z\in \partial \Omega \big|_{y=1}$, the outward normal vector is $n(z) = \begin{pmatrix} 0 \\ 1 \end{pmatrix}$. Therefore, \\
\begin{align}\left\langle n(z), \bar{v}(z) \right\rangle
&= - \beta_2  \leq 0 \qquad \qquad \qquad \forall x \in [0,1] \label{y1_border} \end{align}
\item For each point $z$ along the boundary of $\Omega$ where $x=0$; $z\in \partial \Omega \big|_{x=0}$, the outward normal vector is $n(z) = \begin{pmatrix} -1 \\ 0 \end{pmatrix}$. Therefore, \\
\begin{align}\left\langle n(z), \bar{v}(z) \right\rangle
&= -\alpha_1 \leq 0 \qquad \qquad \qquad \forall y \in [0,1] \label{x0_border} \end{align}
\item For each point $z$ along the boundary of $\Omega$ where $x=1$; $z\in \partial \Omega \big|_{x=1}$, the outward normal vector is $n(z) = \begin{pmatrix} 1 \\ 0 \end{pmatrix}$. Therefore, \\
\begin{align} \left\langle n(z), \bar{v}(z) \right\rangle
&= \alpha_1 - \beta_1 \frac{y^{n_1}}{K_1^{n_1} + y^{n_1}} \label{x1_border} \end{align}
\end{itemize}

Along the boundary $\partial \Omega |_{x=1}$, Equation (\ref{x1_border}) is less than or equal to zero only when $y \geq \hat{y}$, with \[ \hat{y} = \frac{K_1}{\big{(} \frac{\beta_1}{\alpha_1} -1 \big{)}^{\frac{1}{n_1}}}. \]

This implies that trajectories on this boundary enter the feasible set for sufficiently large values of $y$. A complication arises for $y < \hat{y}$, which imposes a restriction on the domain. Since the system cannot be explicitly solved, determining this domain precisely is challenging.

According to the \nameref{thm: pl}, since $g(x,y)$ and $f(x,y)$ are locally Lipschitz, the uniqueness of solutions guarantees that the system's trajectories remain distinct and non-intersecting. Additionally, along the solutions of the dynamical system, the \nameref{thm: tangent} for invariant sets is satisfied since the inner product of the tangent and the norm is always equal to zero. Consequently, the domain is further restricted by the solution passing through the point $(1, \hat y)$.

When considering the \nameref{thm: tangent} at a corner of an invariant set, the challenge is that the corners represent singularities in the geometry of the boundary. The notion of an outward normal vector is well-defined for smooth boundaries, however, for a corner or any non-smooth point on the boundary, the outward normal vector is not unique. Instead, the concept of generalized outward normal vectors is used to describe the possible directions of the outward normals at such points.

For a corner (or any point where multiple boundaries meet), each contributing boundary at the corner will have its own outward normal vector. As a result, the generalized outward normal cone is defined to encompass all possible outward normals at that point. Specifically, at a corner, the generalized outward normal vector can be any convex combination of the outward normal vectors of the boundaries that meet at that corner.

Consider the corner point \( z = (0,0)\). The boundary at this point is defined by two intersecting lines, the vertical boundary $\partial \Omega |_{x=0}$ with outward normal $ \begin{pmatrix} -1 \\ 0 \end{pmatrix}$, and the horizontal boundary $\partial \Omega |_{y=0}$ with outward normal $ \begin{pmatrix} 0 \\ -1 \end{pmatrix}$.

The generalized outward normal cone at the corner \(z=(0,0)\) is the convex cone generated by these two vectors;
\[
N_D(z) = \text{cone}\left\{ \begin{pmatrix} -1 \\ 0 \end{pmatrix}, \begin{pmatrix} 0 \\ -1 \end{pmatrix} \right\}.
\]

In terms of the generalized outward normal cone, we can write this as:
\[
N_D(z) = \{ (n_x, n_y) \in \mathbb{R}^2 \mid n_x \leq 0 \text{ and } n_y \leq 0 \}.
\]
This normal cone defines the set of all valid outward directions from the corner. This generalization of outward normals helps handle non-smooth domains where classical notions of normal vectors do not apply due to corners or other irregularities.

If $n(z) \in N_D(z)$, then \begin{align} \left\langle n(z), \bar{v}(z) \right\rangle = \begin{pmatrix} n_x \\ n_y \end{pmatrix} \cdot \begin{pmatrix} \alpha_1 \\ 0 \end{pmatrix}
&= n_x\alpha_1 \leq 0 \label{pt_border} \end{align}

This concludes that the \nameref{thm: tangent} for invariance is satisfied at the point $z=(0,0)$. At the remaining non-smooth points on the domain, a similar analysis can be conducted to show that the \nameref{thm: tangent} is satisfied.

\begin{figure}[H]
\centering
\includegraphics[scale=0.6]{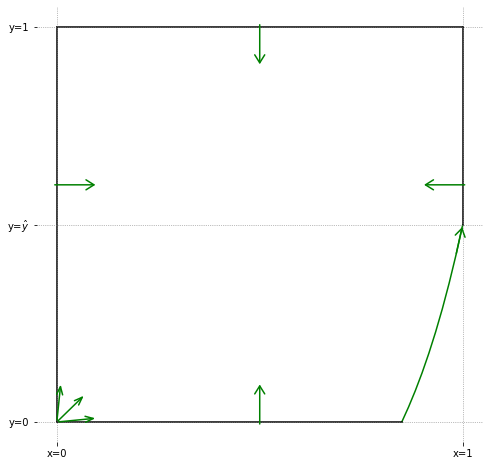}
\caption{Following the discussion of this section, the convex domain of the system is the compact set bounded by the lines $y=0, y=1,x=0,x=1$ and the solution passing through the point $(1,\hat{y})$. Trajectories are bounded within the convex domain illustrated.}
\label{fig: domain}
\end{figure}

By the relationships in (\ref{y0_border})-(\ref{pt_border}), and since $g(x,y)$ and $f(x,y)$ are locally Lipschitz, the \nameref{thm: invar} indicates that the domain is flow-invariant.
Consequently, the convex domain is the compact set bounded by the lines $y=0, y=1,x=0,x=1$ and the solution passing through the point $(1,\hat{y})$, as illustrated in Figure \ref{fig: domain}.

Furthermore, since $g(x,y)$ and $f(x,y)$ are locally Lipschitz, the \nameref{thm: le} Theorem is satisfied. By \nameref{thm: dyn_sys}, this system is indeed a dynamical system.

\subsection{Equilibria and Stability Analysis of the Dynamical System Model Presented in Section \ref{subsec: model}}\label{subsec: equilibria}

\begin{figure}[H]
\centering
\includegraphics[scale=0.65]{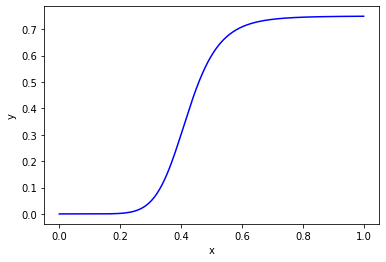}
\caption{Values of [CDK1], or $x$, and [APC], or $y$, for which $f(x,y) = 0$.}
\label{fig: f=0}
\end{figure}

In the setting of the dynamical system demonstrated in Section \ref{subsec: domain}, the system has an equilibrium point when both $g(x,y)= 0 $ and $f(x,y) = 0$ hold simultaneously. The curve for which $f(x,y)= 0$ has the explicit formula given in Equation (\ref{Eq: 5}), and, by using the parameter values previously mentioned: $\alpha_1 = 0.02, \alpha_2 = 3, \alpha_3 = 3, \beta_1 = 3,\beta_2 = 1, K_1 = 0.5, K_2 = 0.5, K_3 = 0.5, n_1 = 8, n_2 = 8, $ and $n_3 = 8$, the resulting curve is illustrated in Figure \ref{fig: f=0}.

\begin{align}
y(x) = \frac{\alpha_2  x^{n_2}}{(\alpha_2  + \beta_2 )x^{n_2} + K_2^{n_2} \beta_2} \label{Eq: 5}
\end{align}

To determine the equilibria of the dynamic system, one must evaluate the values for which $g(x,y)=0$ as well. This task is challenging, as it cannot be solved explicitly.

It is noted that $g(0,y(0))=\alpha_1$. The rate constant of cyclin synthesis $\alpha_1$ is a positive quantity, as it represents the cyclins which are continuously produced in prior phases of the cell cycle to drive cell cycle progression. Therefore, for the system to make physical sense, it is required that $\alpha_1 > 0$. Furthermore, we can express \[g(1,y(1))=\alpha_1 - \beta_1 \frac{y(1)^{n_1}}{K_1^{n_1} + y(1)^{n_1}}.\] By the intermediate value theorem, if $g(1,y(1)) \leq 0$, then there must exist at least one value $x_0$, with $0 \leq x_0 \leq 1$ such that $g(x_0, y(x_0)) = 0$. This establishes a condition that guarantees the existence of an equilibrium point. The condition is given by:

\begin{align} \alpha_1 - \beta_1 \frac{[y(1)]^{n_1}}{K_1^{n_1} + [y(1)]^{n_1}} \leq 0.  \label{ep_existance} \end{align}

If $y(1)$ is in the domain of the system, then following the discussion in Section \ref{subsec: domain}, $y(1) \geq \hat y$. Notably, condition (\ref{ep_existance}) is equivalent to $y(1) \geq \hat y$ (Appendix \nameref{A4}).

\begin{figure}[H]
\centering
\includegraphics[scale=0.65]{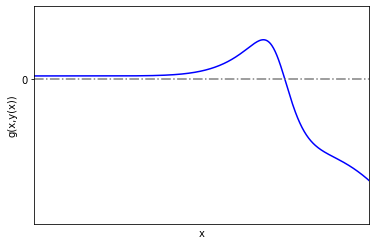}
\caption{The function $g(x,y(x))$, where $y(x)$ is as specified in Equation \ref{Eq: 5}. The roots of the curve displayed are thus used to determine the equilibria of the system. }
\label{fig: g=0}
\end{figure}

Numerical approximation methods were implemented to estimate the equilibria for the given parameters. The values of \( x \) and \( y \) along the curve described in Equation (\ref{Eq: 5}) were substituted into Equation (\ref{Eq: 3}), and plotted in Figure \ref{fig: g=0}. As illustrated in Figure \ref{fig: g=0}, the curve crosses the $x$-axis at only one value of \( x \) in the domain, which identifies the system’s equilibrium point.

\begin{figure}[H]
\centering
\includegraphics[scale=0.55]{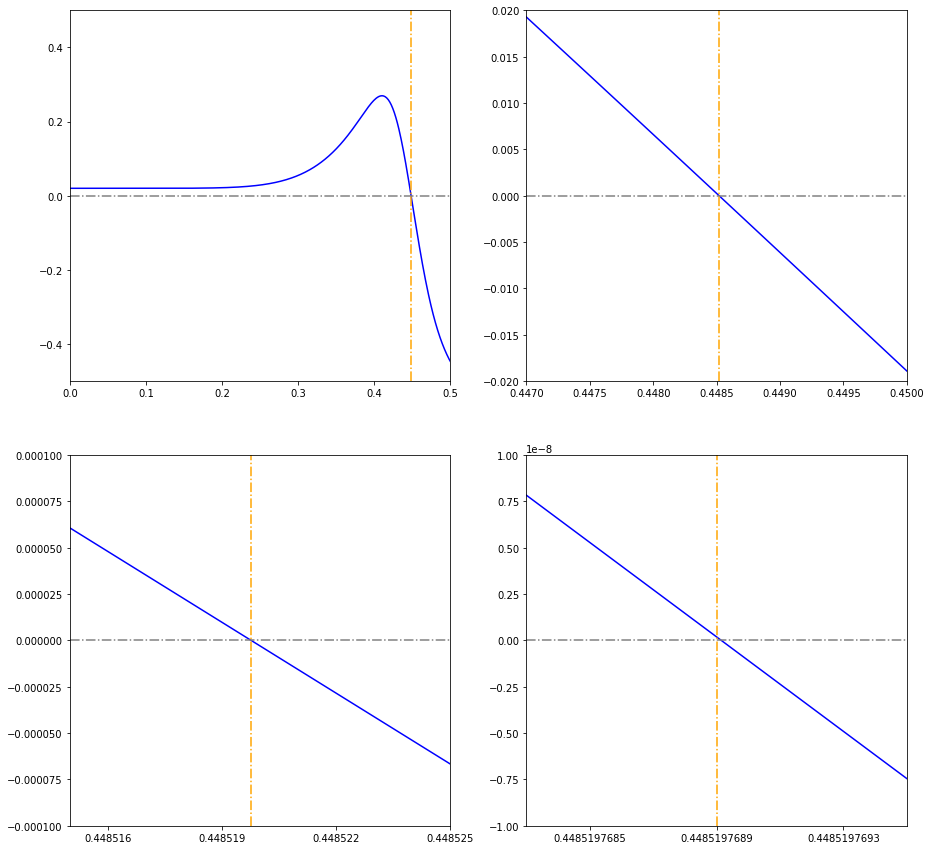}
\caption{Approximate value of $x$ that determines the root of the curve displayed in Figure \ref{fig: g=0}.}
\label{fig: root}
\end{figure}

The method used to approximate the equilibrium point was to determine the root of the curve displayed in Figure \ref{fig: g=0}, approximate the value of $x$, and insert this value into Equation (\ref{Eq: 5}) to approximate the value of $y$ at the equilibrium point. This process is depicted in Figure \ref{fig: root}, where the equilibrium point (accurate to ten significant figures) is: \[ [CDK1] = 0.4485197689 \text{ and } [APC] = 0.4698435624. \]

\begin{figure}[H]
\centering
\includegraphics[scale=0.65]{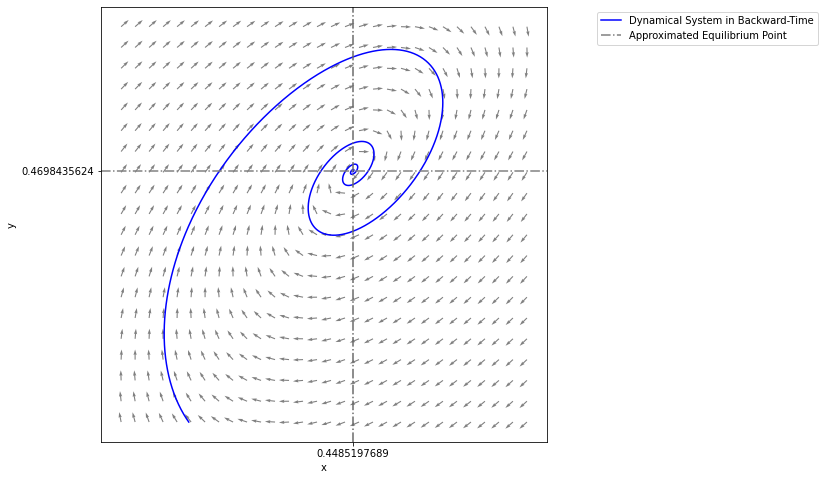}
\caption{Phase portrait and solution curve for the backward-time dynamical system. The equilibrium point is attracting/stable in this system, indicating that the equilibrium point of the original system is repelling/unstable.}
\label{fig: back-time}
\end{figure}

To further investigate the validity of this approximation, and to analyse the stability of the equilibrium point, the dynamical system was solved in negative time (backward time). The phase portrait of the backward-time system is displayed in Figure \ref{fig: back-time}. The equilibrium point is attractive in backward time, and so it can be concluded that this equilibrium point is repelling for the original system.

\begin{figure}[H]
\centering
\includegraphics[scale=0.65]{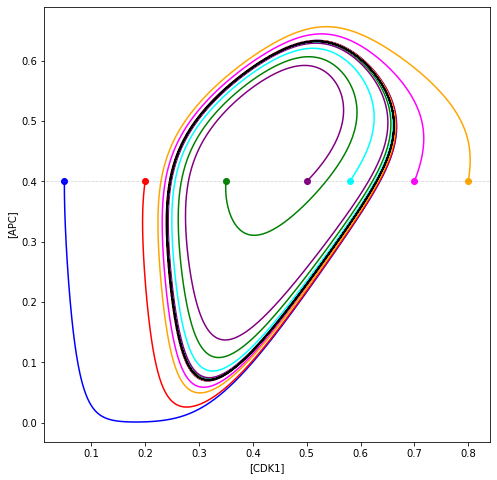}
\caption{Solution curves for the system under various initial conditions. There is one unique stable limit cycle for this dynamical system.}
\label{fig: unique_LC}
\end{figure}

Theoretically, in Section \ref{subsec: domain}, it was proved that the domain of the system is bounded, and in the current section, it has been numerically determined that there is only one repelling equilibrium point for the given parameter values.
By the \nameref{thm: pb}, this indicates that a limit cycle exists for the system. Moreover, Figure \ref{fig: unique_LC} illustrates that this limit cycle is stable and unique, by demonstrating that all trajectories for various initial conditions are attracted to the same limit cycle.

\subsection{Simplified Piece-Wise Smooth System}\label{subsec: aux}

Despite its advantages, mathematical modeling faces challenges. Biological systems are inherently complex, and simplifying assumptions are often necessary, which can reduce the precision of these models. Although dynamical systems theory typically assumes that the system evolution is defined by a smooth function of its arguments, as discussed in Section \ref{subsec: theory}, the study of piecewise smooth dynamical systems is crucial for theoretical advancements and practical applications.
Such problems are all characterized by functions that are piecewise-smooth but are event-driven leading to a loss of smoothness at specific, instantaneous occurrences, such as when a switch is activated/deactivated. These systems not only exhibit intriguing dynamics but also offer valuable practical applications and a rich underlying mathematical structure \cite{piece_smooth}.

\begin{figure}[H]
\centering
\includegraphics[scale=0.65]{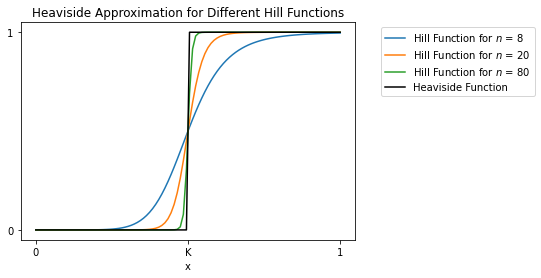}
\caption{Hill functions approximate the Heaviside function as $n \to \infty$.}
\label{fig: heaviside}
\end{figure}

To simplify the system using a piecewise-continuous approximation, the Hill Functions were approximated using the Heaviside Function
\begin{equation}
H(K, x) =
	\begin{cases}
      		0 & \text{where $x \leq K$}\\
      		1 & \text{otherwise}
    	\end{cases}
\end{equation}
for the values of $K_1, K_2, K_3$. To show the validity of the approximation, Figure \ref{fig: heaviside} shows that the Hill function is well approximated for large values of $n$, where $K = 0.5$.

Keeping the naming convention discussed in Section \ref{subsec: domain}, and assuming the simplifying condition $K_3 = K_2$, the system then becomes:

\begin{align}
\frac{dx}{dt} = g(x,y) &:= \alpha_1 - \beta_1 x H(K_1, y) + \alpha_3 (1-x) H(K_2, x) \label{Eq: 7}\\
\frac{dy}{dt} = f(x,y) &:=  \alpha_2 (1 - y) H(K_2, x) - \beta_2 y \label{Eq: 8}
\end{align}

This piecewise continuous system simplifies the analysis of the model by breaking the state space into regions where the dynamics can be explicitly solved in each region, making it easier to study and interpret. This is particularly insightful since the original system cannot be explicitly solved, as discussed in Section \ref{subsec: domain}. Despite this simplification, the essential interactions between variables remain intact, ensuring that the fundamental behavior of the system is preserved. Key structural features of the dynamical system, such as equilibrium points and periodic orbits, are preserved in the approximation. The piecewise continuous system ensures that the global behavior of the system remains qualitatively similar.

The system is split into four regions, and the following holds within each region:

\begin{itemize}
\item In the first region, $0 \leq x \leq K_2$ and $0 \leq y \leq K_1$:
\begin{itemize}
\item[$\circ$] The dynamical system in this region becomes: \\
$\frac{dx}{dt} = g(x,y) := \alpha_1$ \\
$\frac{dy}{dt} = f(x,y) := - \beta_2 y$ \\
\item[$\circ$] This system has the following general solutions: \\
$x(t) = \alpha_1 t + c_1$ \\
$y(t) = c_2 e^{-\beta_2 t} $ \\
\item[$\circ$] The system has one asymptotically stable equilibrium point, being: \\
$x^* = \infty$ \\
$y^* = 0$ \\
\end{itemize}

\item In the second region, $K_2 \leq x \leq 1$ and $0 \leq y \leq K_1$:
\begin{itemize}
\item[$\circ$] The dynamical system in this region becomes: \\
$\frac{dx}{dt} = g(x,y) := \alpha_1 + \alpha_3 (1-x)$ \\
$\frac{dy}{dt} = f(x,y) :=  \alpha_2 (1 - y) - \beta_2 y$ \\
\item[$\circ$] This system has the following general solutions: \\
$x(t) =  \frac{\alpha_1 + \alpha_3}{\alpha_3} + c_3 e^{-\alpha_3 t}$ \\
$y(t) =  \frac{\alpha_2}{\alpha_2 + \beta_2} + c_4 e^{- (\alpha_2 + \beta_2) t}$ \\
\item[$\circ$] The system has one asymptotically stable equilibrium point, being: \\
$x^* =  \frac{\alpha_1 + \alpha_3}{\alpha_3} $ \\
$y^* = \frac{\alpha_2}{\alpha_2 + \beta_2} $ \\
\end{itemize}

\item In the third region, $K_2 \leq x \leq 1$ and $K_1 \leq y \leq 1$:
\begin{itemize}
\item[$\circ$] The dynamical system in this region becomes: \\
$\frac{dx}{dt} = g(x,y) := \alpha_1 - \beta_1 x + \alpha_3 (1-x)$ \\
$\frac{dy}{dt} = f(x,y) :=  \alpha_2 (1 - y) - \beta_2 y$ \\
\item[$\circ$] This system has the following general solutions: \\
$x(t) =  \frac{\alpha_1 + \alpha_3}{\alpha_3 + \beta_1} + c_5 e^{-(\alpha_3+\beta_1) t}$ \\
$y(t) =  \frac{\alpha_2}{\alpha_2 + \beta_2} + c_6 e^{- (\alpha_2 + \beta_2) t}$ \\
\item[$\circ$] The system has one asymptotically stable equilibrium point, being: \\
$x^* =  \frac{\alpha_1 + \alpha_3}{\alpha_3 + \beta_1} $ \\
$y^* =\frac{\alpha_2}{\alpha_2 + \beta_2} $ \\
\end{itemize}

\item In the fourth region, $0 \leq x \leq K_2$ and $K_1 \leq y \leq 1$:
\begin{itemize}
\item[$\circ$] The dynamical system in this region becomes: \\
$\frac{dx}{dt} = g(x,y) := \alpha_1 - \beta_1 x $\\
$\frac{dy}{dt} = f(x,y) :=  - \beta_2 y$ \\
\item[$\circ$] This system has the following general solutions: \\
$x(t) = \frac{\alpha_1}{\beta_1} + c_7 e^{-\beta_1t}$ \\
$y(t) = c_8 e^{-\beta_2 t} $ \\
\item[$\circ$] The system has one asymptotically stable equilibrium point, being: \\
$x^* = \frac{\alpha_1}{\beta_1}$ \\
$y^* = 0$ \\
\end{itemize}
\end{itemize}

where the constants $c_i$ for $i \in {1,2,...,8}$ are determined depending on the initial conditions for each region.

In the context of a piecewise smooth dynamical system, the key focus is on ensuring that the trajectories of the system transition smoothly across the different regions of its phase space.

Let $B_{i,j}$ denote the boundary between region $i$ and region $j$. Trajectories should pass from one region to the next without a jump if smoothness is to be maintained. Therefore, a transition condition must be met to ensure that the system moves from one region to another without any discontinuities. Specifically, the initial condition in region $j$ is the boundary point of the trajectory in the previous region $i$ such that the trajectory crosses $B_{i,j}$.

In order for the system to have a continuous trajectory across regions, the following conditions must hold:

If \( t = t_1 \) is the time at which the system leaves the first region and enters the second region, then the state at the boundary \( (K_2, y(t_1))  \) must be the initial condition for the second region. Similar relationships must hold for the subsequent regions.

Additionally, the asymptotically stable equilibria of each region must lie outside of the region to ensure that the trajectory leaves the region, where the dynamics of the new region are then followed.

It is easy to verify that the asymptotically stable equilibrium point of the first region $( \infty, 0)$ guarantees that the trajectories within this region will indeed enter the second region, with trajectories crossing the boundary $x=K_2$ and $y<K_1$.

Since the equilibrium point leaves the domain in the second region ($x^*>1$), further mathematics is required. The general solution in region 2, explicitly expressed as $y$ as a function of $x$ is given by (Appendix \nameref{A3}):

\begin{align}
y(x) &= \frac{1}{\alpha_2 + \beta_2} \bigg( \alpha_2 - C( \alpha_1 + \alpha_3 -\alpha_3 x )^{\frac{{\alpha_2+\beta_2}}{\alpha_3}} \bigg) \label{Eq: y(x)_region2}
\end{align}

Obtain the trajectory $y(x)$ with initial condition $(x_0, y_0) = (K_2, 0)$, from Equation (\ref{Eq: y(x)_region2}) (Appendix \nameref{A3})

\[ y(x) = \frac{\alpha_2}{\alpha_2 + \beta_2} \bigg( 1 - \bigg( \frac{\alpha_1 + \alpha_3 -\alpha_3 x} { \alpha_1 + \alpha_3 -\alpha_3 K_2} \bigg)^{\frac{{\alpha_2+\beta_2}}{\alpha_3}} \bigg). \]

For the solution to remain in the feasible set and cross the correct border, it must be true that when $y(\hat x)=K_1$, $K_2 \leq \hat x \leq 1$. Consider this value for $\hat x$, which results in:

\begin{align}
\hat x &= 1 + \frac{\alpha_1}{\alpha_3} - \bigg( \frac{\alpha_1}{\alpha_3}  + 1- K_2 \bigg) \bigg (\frac{-\beta_2}{\alpha_2} K_1 + 1 - K_1 \bigg)^{\frac{\alpha_3}{\alpha_2+\beta_2}}. \nonumber
\end{align}

For this value of $\hat x$, if $K_2 \leq \hat x \leq 1$, then the solution intersects the line $y=K_1$, and trajectories enter region three.

In region three, the equilibrium point is

\[ x^* =  \frac{\alpha_1 + \alpha_3}{\alpha_3 + \beta_1}, y^* =\frac{\alpha_2}{\alpha_2 + \beta_2}. \]

To ensure that the transition condition is met, the following relationships must be maintained:
\begin{itemize}
\item $0 <  x^* < K_2$,
\item $K_1 < y^* < 1 $.
\end{itemize}

Or, more explicitly, since $y^* = \frac{\alpha_2}{\alpha_2 + \beta_2}  <1$ always, the following conditions on the parameters guarantee that the trajectories enter the fourth region:
\begin{itemize}
\item $0 <  \frac{\alpha_1 + \alpha_3}{K_2 (\alpha_3 + \beta_1)} < 1$,
\item $1 < \frac{\alpha_2}{K_1 (\alpha_2 + \beta_2)} $.
\end{itemize}

It is again easy to verify that the equilibrium point of the fourth region \[ x^* = \frac{\alpha_1}{\beta_1}, y^* = 0 \] guarantees that the trajectories within this region will indeed cross the line $y=K_1$. One further condition on the $x$ coordinate is required for the equilibrium point to remain within the first region, being $0 < x^* < K_2$, or more explicitly, since $0 < x^*$ always, the parameters must maintain the following relationship:
\[ \frac{\alpha_1}{K_2 \beta_1} < 1. \]

\begin{figure}[H]
\centering
\includegraphics[scale=0.65]{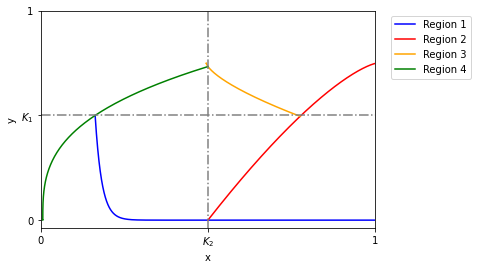}
\caption{Illustration of the auxiliary system in the four different regions using the Heaviside Approximation, which demonstrates that the transition conditions are adhered to.}
\label{split-approx}
\end{figure}

The above discussion is illustrated in Figure \ref{split-approx}. It is clear that by using this auxiliary system, one can easily see the mechanism driving the oscillations of the dynamical system. The system is still oscillatory for $n \to \infty$, and using this auxiliary system enables us to easily determine the explicit solutions within each region, while being able to derive explicit conditions on the parameters to ensure oscillatory behaviour.

Similar to the investigation performed in Section \ref{subsec: domain}, to determine whether the system's trajectories remain within the feasible set described in (\ref{border}), an analysis is conducted on the boundary $\partial \Omega$ using the \nameref{thm: tangent} (\ref{tangent}). Let $\bar{v} = \begin{pmatrix} \frac{dx}{dt} \\ \\
\frac{dy}{dt}\end{pmatrix} $. Then,

\begin{itemize}
\item In the first region, $0 \leq x \leq K_2$ and $0 \leq y \leq K_1$:
\begin{itemize}
\item [$\circ$]  If $z\in \partial \Omega_1 |_{x=0}$, then $n(z) = \begin{pmatrix} -1 \\ 0 \end{pmatrix}$. Therefore, \\
\begin{align}
\left\langle n(z), \bar{v}(z) \right\rangle
&= -\alpha_1 \leq 0 \qquad \forall y \in [0,K_1] \nonumber \end{align}

\item [$\circ$] If $z\in \partial \Omega_1 |_{y=0}$, then $n(z) = \begin{pmatrix} 0 \\ -1 \end{pmatrix}$. Therefore, \\
\begin{align}
\left\langle n(z), \bar{v}(z) \right\rangle
&= 0 \qquad \forall x \in [0,K_2] \nonumber \end{align}
\end{itemize}

\item In the second region,$K_2 \leq x \leq 1$ and $0 \leq y \leq K_1$:
\begin{itemize}
\item [$\circ$]  If $z\in \partial \Omega_2 |_{x=1}$, then $n(z) = \begin{pmatrix} 1 \\ 0 \end{pmatrix}$. Therefore, \\
\begin{align}
\left\langle n(z), \bar{v}(z) \right\rangle
&= \alpha_1 \geq 0 \qquad \forall y \in [0,K_1] \nonumber \end{align}

\item [$\circ$] If $z\in \partial \Omega_2 |_{y=0}$, then $n(z) = \begin{pmatrix} 0 \\ -1 \end{pmatrix}$. Therefore, \\
\begin{align}
\left\langle n(z), \bar{v}(z) \right\rangle
&= -\alpha_2 \leq 0 \qquad \forall x \in [K_2, 1] \nonumber \end{align}
\end{itemize}

\item In the third region, $K_2 \leq x \leq 1$ and $K_1 \leq y \leq 1$:
\begin{itemize}
\item [$\circ$]  If $z\in \partial \Omega_3 |_{x=1}$, then $n(z) = \begin{pmatrix} 1 \\ 0 \end{pmatrix}$. Therefore, \\
\begin{align}
\left\langle n(z), \bar{v}(z) \right\rangle
&=  \alpha_1 - \beta_1 \leq 0 \text{ provided that }  \alpha_1 \leq \beta_1  \qquad \forall y \in [K_1,1] \nonumber \end{align}

\item [$\circ$] If $z\in \partial \Omega_3 |_{y=1}$, then $n(z) = \begin{pmatrix} 0 \\ 1 \end{pmatrix}$. Therefore, \\
\begin{align}
\left\langle n(z), \bar{v}(z) \right\rangle
&= - \beta_2  \leq 0 \qquad \forall x \in [K_2, 1] \nonumber \end{align}
\end{itemize}

\item In the fourth region, $0 \leq x \leq K_2$ and $K_1 \leq y \leq 1$:
\begin{itemize}
\item [$\circ$]  If $z\in \partial \Omega_4 |_{x=0}$, then $n(z) = \begin{pmatrix} -1 \\ 0 \end{pmatrix}$. Therefore, \\
\begin{align}
\left\langle n(z), \bar{v}(z) \right\rangle
&= -\alpha_1 \leq 0 \qquad \forall y \in [K_1,1] \nonumber \end{align}

\item [$\circ$] If $z\in \partial \Omega_4 |_{y=1}$, then $n(z) = \begin{pmatrix} 0 \\ 1 \end{pmatrix}$. Therefore, \\
\begin{align}
\left\langle n(z), \bar{v}(z) \right\rangle
&= -\beta_2 \leq 0 \qquad \forall x \in [0,K_2] \nonumber \end{align}
\end{itemize}

\end{itemize}

\begin{figure}[H]
\centering
\includegraphics[scale=0.6]{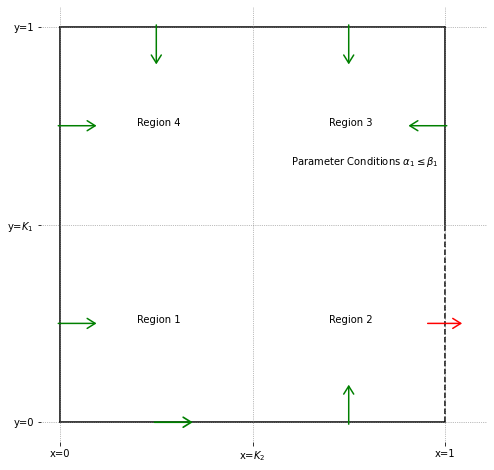}
\caption{Assuming the condition $\alpha_1 \leq \beta_1$ holds, trajectories are directed inwards on all borders, except along the boundary $\partial \Omega_2 |_{x=1}$ in region 2.}
\label{fig: aux_border}
\end{figure}

The above discussion is illustrated in Figure \ref{fig: aux_border}.

Similar to before, along the border $\partial \Omega_2 |_{x=1}$  in the second region, trajectories are not restricted to the feasible set. Hence, a thorough analysis is performed to find the conditions on the parameters that guarantee a closed domain.

From Equation (\ref{Eq: y(x)_region2}) with the initial condition $(x_0, y_0) =(1, K_1)$, the corresponding trajectory is given by (Appendix \nameref{A2}):

\begin{align}
y(x) &= \frac{K_1(\alpha_2-\beta_2)}{\alpha_2 + \beta_2} \bigg( 1 + \frac{\alpha_3}{\alpha_1} (1 - x) \bigg)^{\frac{{\alpha_2+\beta_2}}{\alpha_3}}. \nonumber
\end{align}

In order to determine the restriction on the parameters, consider $\hat x$ such that $y( \hat x)=0$, which results in (Appendix \nameref{A2}):

\begin{align}
\hat x &= 1 - \frac{\alpha_1}{\alpha_3} \bigg( \frac{\alpha_2}{\alpha_2  - K_1 (\alpha_2 + \beta_2)} \bigg) ^{\frac{\alpha_3}{\alpha_2+\beta_2}} + \frac{\alpha_1}{\alpha_3}  \nonumber
\end{align}

According to the \nameref{thm: pl}, the uniqueness of solutions guarantees that the system’s trajectories remain distinct and non-intersecting. Consequently, the exact solution that results from the initial condition $(x_0, y_0) =(1, K_1)$ contributes to the convex domain within which the solutions of the system are bounded.

As a consequence, the condition on the parameters for the domain to be convex and closed is to have \[ K_2 \leq \hat x \leq 1. \]

\begin{figure}[H]
\centering
\includegraphics[scale=0.6]{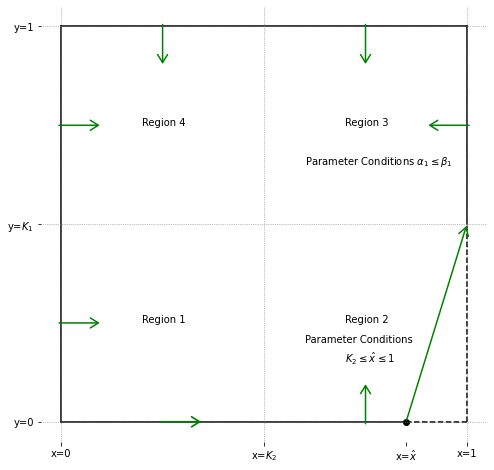}
\caption{Following the discussion of this section, the convex domain of the system is the compact set bounded by the lines $y = 0, y = 1, x = 0, x = 1$ and the solution passing through the point $(\hat x, 0)$. Trajectories are bounded within the convex domain illustrated.}
\label{fig: new_aux_border}
\end{figure}

\begin{figure}[H]
\centering
\includegraphics[scale=0.65]{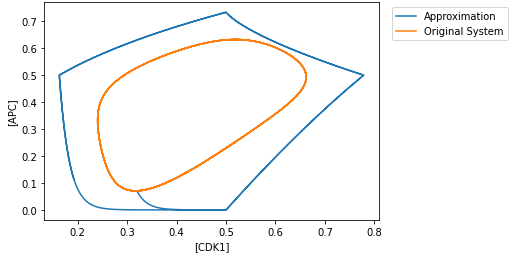}
\caption{Solutions of the auxiliary system and the original system with $n_1=n_2=n_3=8$ and all other parameters remaining the same.}
\label{poor-approx}
\end{figure}

\begin{figure}[H]
\centering
\includegraphics[scale=0.65]{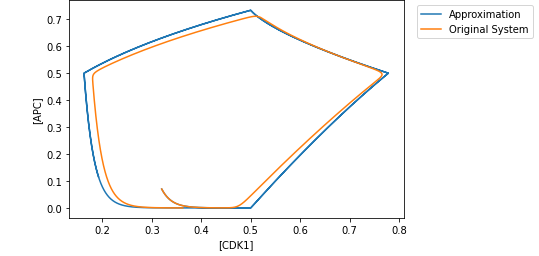}
\caption{Solutions of the auxiliary system and the original system with $n_1=n_2=n_3=80$ and all other parameters remaining the same.}
\label{good-approx}
\end{figure}

\subsection{Impacts of the Parameter $\alpha_1$}\label{subsec: alpha1}

\begin{figure}[H]
\centering
\includegraphics[scale=0.55]{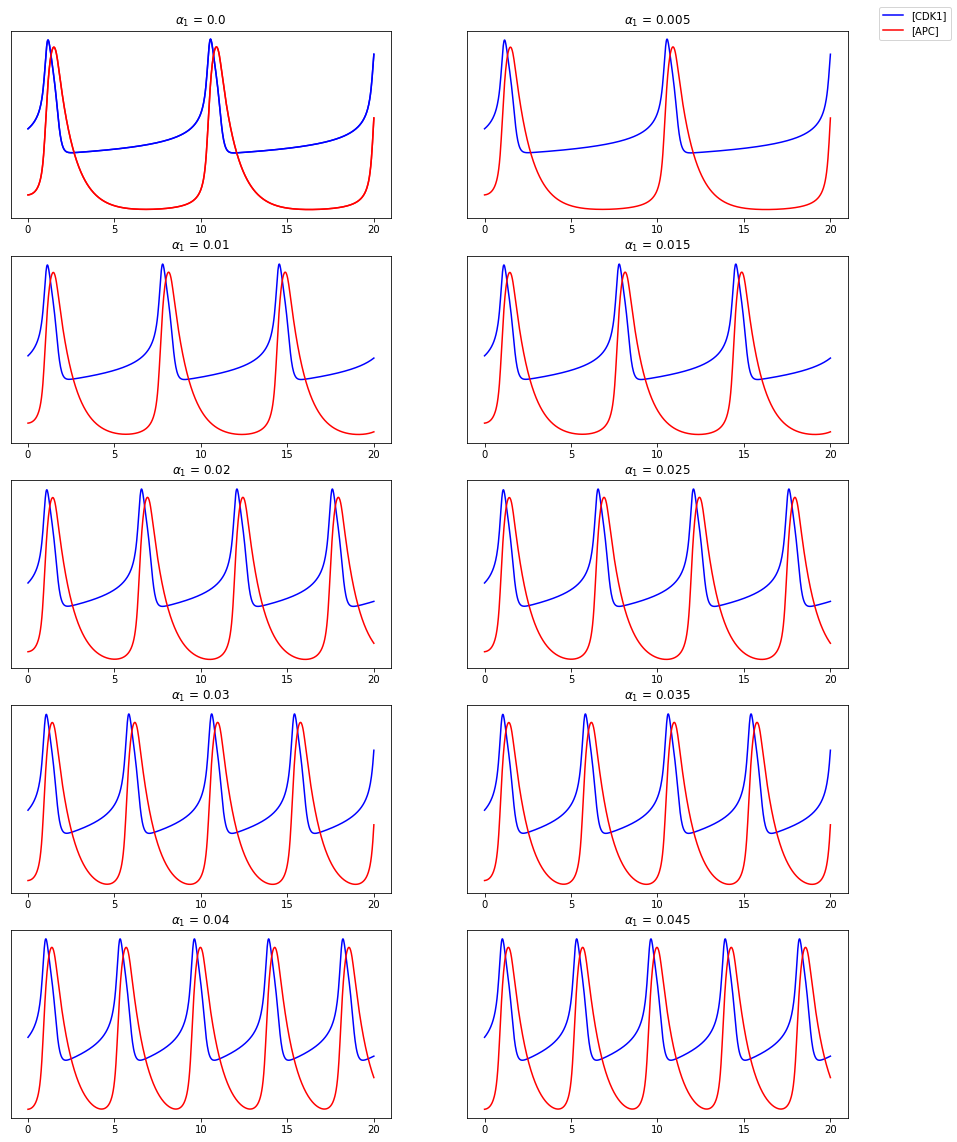}
\caption{Plots showing the solution curves for different values of $\alpha_1$. As $\alpha_1$ increases, the period of the solutions decrease. }
\label{fig: a1_effect}
\end{figure}

To determine the overall effect that the parameter $\alpha_1$, which represents the rate constant of cyclin synthesis, has on the system, the value of $\alpha_1$ was varied and the corresponding solution curves were plotted, shown in Figure \ref{fig: a1_effect}.

It can be observed that an increasing value of $\alpha_1$ leads to a shorter period in the solution curves. In the context of this discussion, a higher value of \( \alpha_1 \) correlates with faster cell proliferation, identifying \( \alpha_1 \) as the control variable in the system. Consequently, to reduce the period of the solution curves and ultimately slow cancer cell progression, one may aim to decrease the rate constant of cyclin synthesis.

\begin{figure}[H]
\centering
\includegraphics[scale=0.5]{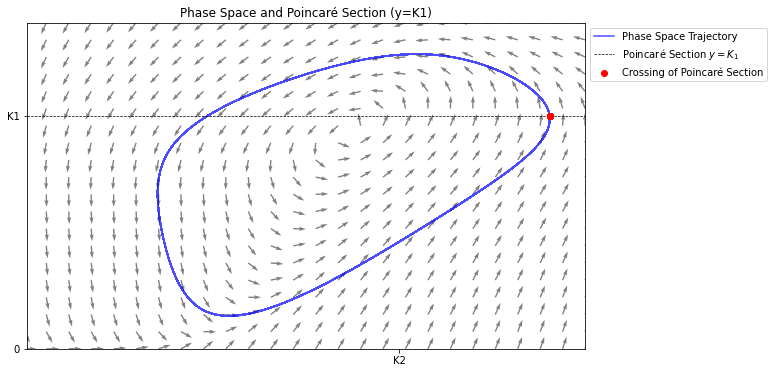}
\caption{Illustration of the method used to determine the period of the limit cycle, calculated for various values of $\alpha_1$.}
\label{fig: poincare}
\end{figure}

\begin{figure}[H]
\centering
\includegraphics[scale=0.5]{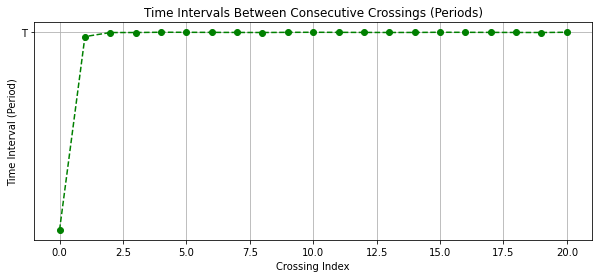}
\caption{As time increases, consecutive crossings occur at approximately the same point. The time between consecutive crossings is approximately constant, and this time closely approximates the period of the limit cycle.}
\label{fig: estimate}
\end{figure}

For systems with a stable limit cycle, as time progresses, the solutions tend toward the limit cycle. A Poincar\'e section is a useful tool for determining the period of the limit cycle. In the two-dimensional case, it works by selecting a line within the phase space of the system, which the trajectory crosses as the system evolves over time. One calculates the times at which the trajectory intersects the Poincar\'e section, noting that as time increases, consecutive crossings occur at approximately the same point. The time between one crossing and the next is approximately constant, and this time closely approximates the period of the limit cycle. This method reduces the complexity of tracking the continuous motion of the system by focusing only on discrete events (being the crossings), making it easier to measure how long it takes for the system to return to the same point in space. Figure \ref{fig: poincare} demonstrates the idea behind determining the period of the limit cycle, and Figure \ref{fig: estimate} emphasises the fact that the times between consecutive crossings of the Poincar\'e section are approximately equal, thus approximating the period of the system well.

To quantitatively describe the effect that $\alpha_1$ has on the period of the solutions, the Poincaré section $y=K_1$ was employed, obtaining points where the trajectory crosses this line from below. By varying the values of $\alpha_1$ and calculating the corresponding periods, a curve fit was performed to illustrate the relationship between the period $T$ of the solutions, and the parameter $\alpha_1$.
To fit data assuming a hyperbolic relationship between the period $T$, and the parameter $\alpha_1$, the $\texttt{curve\_fit}$ function from the $\texttt{scipy.optimize}$ package in Python was used. This function performs non-linear least squares fitting by estimating parameters that best fit the data to the specified function. For this analysis, a hyperbolic function of the form \[ T(\alpha_1; A, B) = \frac{A}{\alpha_1} + B \] was defined. The function $\texttt{curve\_fit}$ was then applied to the calculated data to optimize the parameters \( A \) and \( B \).

\begin{figure}[H]
\centering
\includegraphics[scale=0.6]{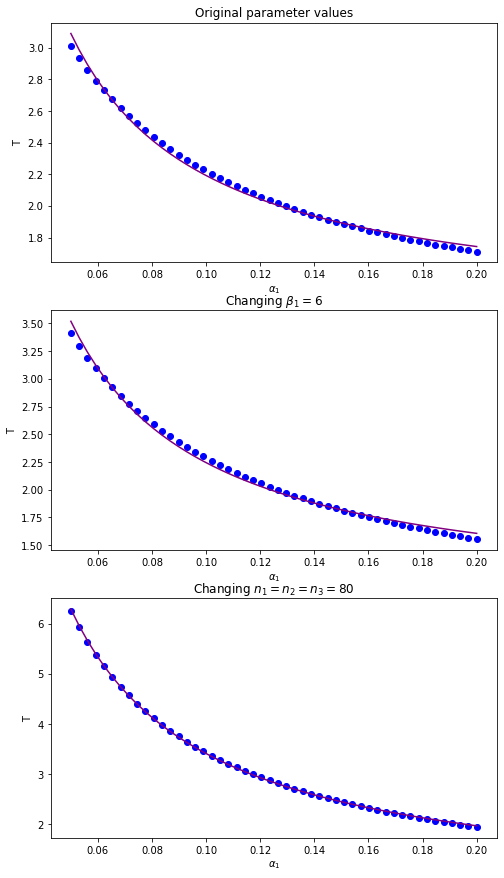}
\caption{Numerically determined relationship between the period of the cell cycle, $T$, and the parameter $\alpha_1$.}
\label{fig: period}
\end{figure}

Figure \ref{fig: period} presents the results of the Poincar\'e section implementation, further emphasising the relationship between the period $T$ and $\alpha_1$. Moreover, the relationship holds for different parameter values (an increased $\beta_1$, or $n_1,n_2,n_3$).

The relationship between the period of the limit cycle and the parameter $\alpha_1$ obtained is thus:

\begin{align}
T &= \frac{A}{\alpha_1} + B \label{eq: T_by_alpha1}
\end{align}

where $A$ and $B$ are parameters that were numerically fitted depending on the other parameter values.

\section{Theoretical Model and Experimental Data Links}\label{sec: theory}

\subsection{Theoretical Cancer Population Model}\label{subsec: pop_model}

In an attempt to mathematically construct the concept of cell viability as a function of time, let $M(t)$ denote the size of a natural population of cancer cells. Suppose that sufficient resources and an optimal environment are provided. A dynamical system for cancer cell population is considered because many different cells may be in different phases in the cell cycle. This would then mean that the population does not grow discretely, but rather continuously.  It is important to note that the average period remains the same even though cells are not dividing at the same time.
Under such conditions, one may assume a constant growth rate model for the population, that is $M(t)$ is a solution of the differential equation
\[ \frac{dM}{dt} = rM \]
where $r > 0$ is the growth rate of the population.

In Section \ref{subsec: alpha1} it was shown that the period of the cell cycle is related to the parameter $\alpha_1$. From equation (\ref{eq: T_by_alpha1}), the period $T$ is a function of cyclin synthesis $\alpha_1$, where it is assumed that $\alpha_1$ remains a constant without treatment.

Theory informs that Zingerone decreases the expression of Cyclin D1 and thus reduces the cyclin expression in the M-phase \cite{zingerone_d1}. As such, the parameter that describes the cyclin expression, $\alpha_1$, has a dependency on the concentration of Zingerone which has penetrated the cell membrane.

Let $c(t,z)$ be the rate of cyclin synthesis under treatment by Zingerone of concentration $z$ at time $t$. Then, $c(t,0) = \alpha_1$, a constant, and the period of the cell cycle under treatment, $\tau$, can be expressed as
\[ \tau(t,z) = \frac{A}{c(t,z)} + B. \]

The mass-action principle results in the following:

\[ \frac{dc(t,z)}{dt} = -\rho c(t,z)z. \]

Here, $c$ is the available active cyclin, which decreases when it interacts with Zingerone of concentration $z$. Note that $c(0,z)=\alpha_1$ to obtain the following:

\[ c(t,z) = \alpha_1 e^{-\rho z t}. \]

Therefore,

\[ \tau(t,z) = \frac{A}{\alpha_1 e^{-\rho z t}} + B. \]

The growth rate of the cancer cell population model is inversely related to the period, as each completed period results in the division of a single cell into two cells. Thus, the growth rate is also dependent on $\alpha_1$, and we have $r(t,z)$. The population model becomes:

\[ \frac{dM(t,z)}{dt} = r(t,z)M(t,z). \]

The doubling time $(\tau_2)$ of a population model is the amount of time it takes for the population to double. In the case of cancer cells, the doubling time is equal to the period of the cell cycle, $\tau_2 = \tau$, because one cell splits into two after the period is complete. Due to this, the exponential growth population model has rate as a function of doubling time given by (Appendix \nameref{A7}) \[ r = \frac{\ln(2)}{\tau_2}. \]

Applying this to the growth rate $r(t,z)$, and assuming a constant death rate $\mu$ of cancer cells,

\begin{align}
r(t,z) &= \frac{\ln(2)}{\tau(t,z)} - \mu \nonumber \\
&=  \frac{\ln(2)}{ \frac{A}{\alpha_1 e^{-\rho z t}} + B} - \mu \nonumber \\
&=  \frac{1}{ \hat A e^{\rho z t} + \hat B} - \mu \nonumber
\end{align}

where \[ \hat A = \frac{A}{\ln(2) \alpha_1}, \text{ and} \]
\[ \hat B = \frac{B}{\ln(2)}. \]

Using the above in the population model for cancer cells, one obtains the following:

\[ M(t,z) = M(0)e^{\int_0^t  \frac{1}{ \hat A e^{\rho z \theta} + \hat B} d\theta - \mu t}. \]

For the untreated population, $z=0$:

\[ M(t, 0) = M(0)e^{ \frac{t}{ \hat A  + \hat B} - \mu t}. \]

Theoretically, the cell viability $CV$ of a treated population is a function of time $t$, and treatment concentration, in this case, $z$. At any given time, it is defined as the ratio of the treated population size over the size of this population if it is left untreated \cite{prof}. The cell viability is the function

\begin{align}
CV(t,z) := \frac{M(t,z)}{M(t,0)} &= e^{\int_0^t  \frac{1}{ \hat A e^{\rho z \theta} + \hat B} d\theta - \mu t - \big[\frac{t}{ \hat A  + \hat B} - \mu t\big]} \nonumber \\
 &= e^{\int_0^t  \frac{1}{ \hat A e^{\rho z \theta} + \hat B} d\theta - \frac{t}{ \hat A  + \hat B} } \nonumber \\
 &= e^{-\frac{1}{\rho \hat B z} \ln \big(\frac{\hat A+\hat Be^{-\rho z t}}{\hat A+\hat B} \big) - \frac{t}{ \hat A  + \hat B} } \label{eq: cv}
\end{align}

where the fraction is expressed as a percentage.

\subsection{Experimental Results and Discussion}\label{subsec: res}

\begin{figure}[H]
\centering
\includegraphics[scale=0.45]{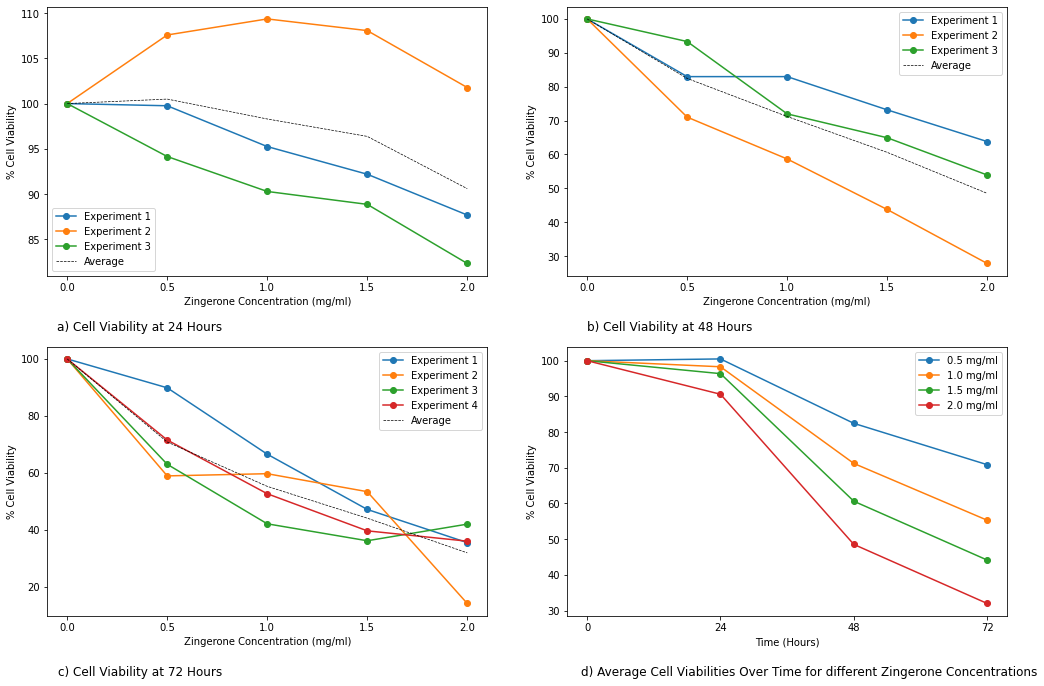}
\caption{Graphs showing experimental data for Zingerone concentrations of 0.5,1.0,1.5, and 2.0 mg/ml at the time points of 24 hours, 48 hours, and 72 hours that were obtained.}
\label{viability}
\end{figure}

Melanoma cells were exposed to Zingerone at concentrations of $0.5, 1.0, 1.5$, and $2.0$ milligrams per milliliter (mg/ml) for $24, 48$, and $72$ hours. The cells were then tested for cell viability. The goal is to measure the effect of Zingerone on cell viability. For each time point, at least three experiments with significance were carried out. Variability across experiments can be observed in Figure \ref{viability} (a) - (c). The trends of cell viability as a function of time and concentration become more evident in the time diagram (d) of the averaged data. The data represented in Figure \ref{viability} can be considered as a set of measurements of the function (\ref{eq: cv}) at times $t = 0, t = 24, t = 48, t = 72$ hours.

A multivariate curve-fitting approach was performed. This procedure estimates the parameters of Equation (\ref{eq: cv}) that best fits the experimental data that was obtained.

The independent variables, time \( t \) and zingerone concentration \( z \), and the dependent variable, cell viability \( CV \), being the experimental data, were organized into arrays suitable for $\texttt{scipy.optimize.curve\_fit}$ in Python. The resulting best-fit function was given by:

\begin{align}
CV(t,z) &= e^{-\frac{1}{1.4006 z} \ln ( 0.20333 + 0.79667 e^{-0.041156 tz} ) - \frac{t}{ 42.716} } \label{eq: cv_fit}
\end{align}

\begin{figure}[H]
\centering
\includegraphics[scale=0.45]{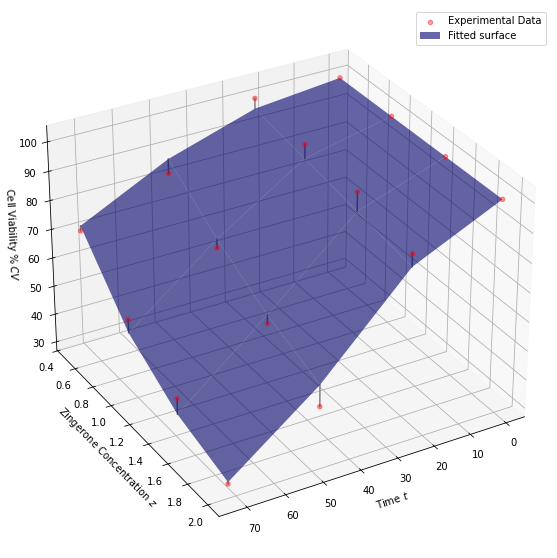}
\caption{3D Curve fit of theoretical model on experimental data for Zingerone concentrations of 0.5,1.0,1.5 and 2.0 mg/ml at the time points of 24 hours, 48 hours, and 72 hours.}
\label{3d_fit}
\end{figure}

\begin{figure}[H]
\centering
\includegraphics[scale=0.45]{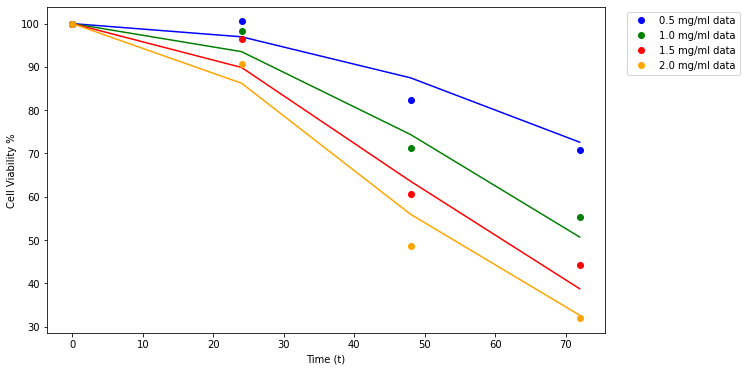}
\caption{2D Curve fit of theoretical model on experimental data for Zingerone concentrations of 0.5,1.0,1.5 and 2.0 mg/ml at the time points of 24 hours, 48 hours, and 72 hours, where the solid lines represent the curve fitted plot at these values.}
\label{viability_fit}
\end{figure}

The experimental data and fitted surface are shown in Figure \ref{3d_fit}, and the results are emphasised in Figure \ref{viability_fit}.

Note that since the data availability is very limited, the performance of the theoretical cell viability model has a lot of room for improvement. Additionally, here, a constant death rate was assumed. However, it is likely that with an increased concentration of Zingerone over a longer time period, cell death might become dependent on these parameters. Overall, this paper shows that the mathematical model describing the interactions between CDK1 and APC in Section \ref{subsec: fullmodel} is a relatively accurate depiction of real cancer cells. This paper has given the model mathematical validity, as well as comparisons with real-world data, securing the assumption that this model is indeed a good fit for cancer cells.

\subsection{IC$_{50}$ curve}

IC\(_{50}\) is a widely used metric for assessing the inhibitory effectiveness of a substance and serves as a standard for comparing the potency of different agents. Specifically, at a given time \( t \), IC\(_{50}\)(t) represents the concentration required to decrease cell viability by 50\% at time $t$ \cite{prof}.

\begin{figure}[H]
\centering
\includegraphics[scale=0.6]{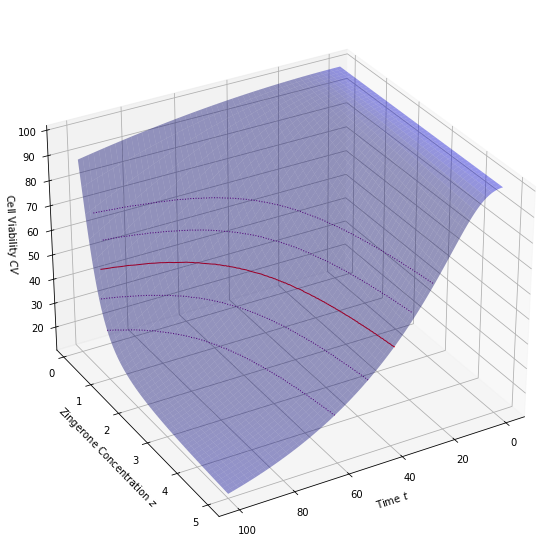}
\caption{ Surface graph of the cell viability $CV$ defined in explicit solution given in Equation \ref{eq: cv_fit} as a function of the time $t$ and the concentration of Zingerone $z$. In this figure, the solid line represents the level curve at 50\%, while the dashed lines (from bottom to top) represent the level curves at 30\%, 40\%, 60\%, and 70\%, respectively.}
\label{3d}
\end{figure}

\begin{figure}[H]
\centering
\includegraphics[scale=0.5]{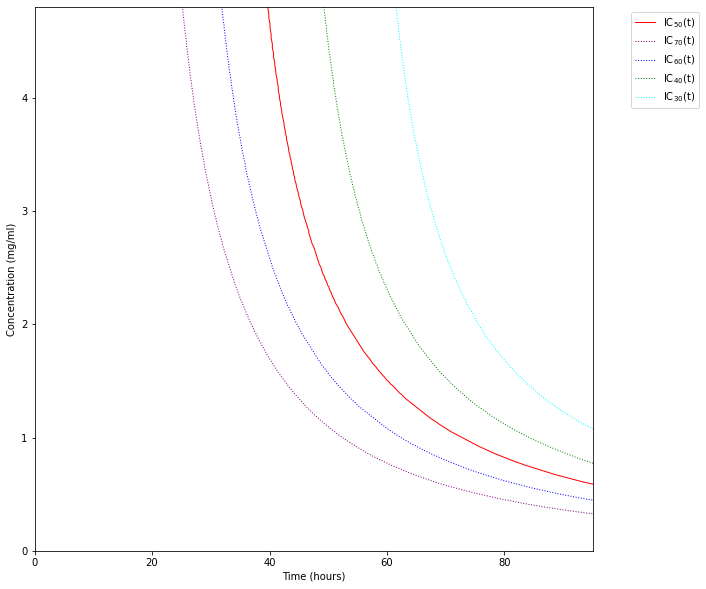}
\caption{Here, the solid line represents IC$_{50}(t)$ as a function of time, while the dashed lines (from left to right) represent IC$_{30}(t)$, IC$_{40}(t)$, IC$_{60}(t)$, IC$_{70}(t)$ as functions of time, respectively.}
\label{2d}
\end{figure}

The graph of the cell viability function \( CV \), as defined in Equation (\ref{eq: cv_fit}), is shown in Figure \ref{3d}. In this figure, the solid line on the surface represents the intersection with the horizontal plane where cell viability is equal to 50\%. This intersection provides a visual representation of the IC\(_{50}\) value, which indicates the concentration of Zingerone that reduces cell viability by half. When plotted against concentration and time, the function yields the graph of IC\(_{50}\)(t), as depicted by the solid line in Figure \ref{2d}. This graph enables the determination of IC\(_{50}\)(t) at any specific point in time, offering valuable insights into how the inhibitory concentration changes over time.

In addition, since the cell viability function is expressed explicitly in Equation (\ref{eq: cv_fit}), we can generate level curves for any given level of cell viability. These level curves represent different cell viabilities and can be constructed at various time points. The dashed lines in Figures \ref{3d} and \ref{2d} illustrate the level curves corresponding to cell viability values of 30\%, 40\%, 60\%, and 70\% as functions of time. By using these curves, we can determine not only the IC\(_{50}\) value for a specific time point, but also identify the concentration required to achieve any desired level of cell viability, offering a more flexible approach to evaluating the effects of the substance at different concentrations and time intervals.

The decreasing trend of IC$_{50}$ over time further emphasises that Zingerone becomes more effective as time progresses.

\section{Conclusion}\label{sec: conc}

This report highlights the critical role of mathematical modeling in advancing our understanding of cancer cell dynamics and offers a framework that links theoretical rigor with biological relevance.

By introducing and thoroughly examining a biologically derived dynamical system for cancer cell cycle behaviour, we successfully established the bounded domain of the system, and identified conditions under which the system exhibits a stable limit cycle,  supporting the hypothesis that certain regulatory mechanisms can lead to oscillatory behavior in cancer cell cycles. Additionally, we obtained a relationship between the period of the cell cycle, and a control parameter, \(\alpha_1\) mathematically. The process of simplifying the original model into a piecewise smooth dynamical system, while preserving its qualitative properties, emphasised that the relationship between the period of the cell cycle and \(\alpha_1\) holds even for the simplified system, thereby enhancing the robustness of our findings across modeling approaches.

The experimental validation using Zingerone as a modulating factor on the cell cycle period further demonstrates the power of combining mathematical modeling with empirical data. By fitting experimental results to our theoretical model, we confirmed that Zingerone effectively reduces the period of the cell cycle, suggesting its role in slowing cancer cell proliferation, which aligns with Zingerone’s proposed therapeutic potential in cancer treatment. This integration of theory and experiment emphasizes how mathematical models can be pivotal in preclinical evaluations of novel therapeutic compounds, providing an efficient, cost-effective approach to predict biological responses before conducting extensive laboratory tests.

This research lays a foundation for future studies exploring mathematical models as predictive tools in oncology. The model presented here can be extended to include additional cellular interactions, environmental factors, or drug interventions, offering an adaptable approach for exploring more complex systems. Moreover, our results encourage further investigation into the relationship between cell cycle period and regulatory parameters like \( \alpha_1 \), which could inspire new therapeutic strategies aimed at controlling cell proliferation through parameter adjustments. Overall, mathematical modeling emerges as an indispensable tool in cancer research, aiming to accelerate the discovery of targeted therapies through insightful, theoretically grounded approaches.

\bibliography{refs}{}
\bibliographystyle{unsrt}

\newpage
\section*{Appendix}\label{sec: Appendix}

\subsubsection*{A1}\label{A5}
Proof that $g(x,y)$ and $f(x,y)$ are Locally Lipschitz as discussed in Section \ref{subsec: domain}:

\begin{align}
\frac{\partial}{\partial x} g(x,y) &:= - \beta_1 \frac{y^{n_1}}{K_1^{n_1} + y^{n_1}} + \alpha_3 \bigg[- \frac{x^{n_3}}{K_3^{n_3} + x^{n_3}} + (1-x) \frac{n_3 K_3^{n_3} x^{n_3-1}}{(K_3^{n_3} + x^{n_3})^2}\bigg] \nonumber\\
\frac{\partial}{\partial y} g(x,y) &:= - \beta_1 x  \frac{n_1 K_1^{n_1} y^{n_1-1}}{(K_1^{n_1} + y^{n_1})^2}  \nonumber\\
\frac{\partial}{\partial x} f(x,y) &:=  \alpha_2 (1 - y)  \frac{n_2 K_2^{n_2} x^{n_2-1}}{(K_2^{n_2} + x^{n_2})^2} \nonumber \\
\frac{\partial}{\partial y} f(x,y) &:=  - \alpha_2 \frac{x^{n_2}}{K_2^{n_2} + x^{n_2}} - \beta_2 \nonumber
\end{align}

Since the above equations all define continuous functions on $D:= \{\mathbb{R}^2| x \geq 0, y\geq 0\}$, the conditions of Proposition \ref{prop: ll} are satisfied, and thus $g(x,y)$ and $f(x,y)$ are Locally Lipschitz on $D$.

\subsubsection*{A2}\label{A4}
Assuming $y(1)$ from the discussion in Section \ref{subsec: equilibria} is in the domain of the system, then from Section \ref{subsec: domain}, we have that $y(1) \geq \hat y$. This means:

\begin{align}
y(1) &\geq \frac{K_1}{\bigg( \frac{\beta_1}{\alpha_1} - 1 \bigg)^{\frac{1}{n_1}}} \nonumber \\
[y(1)]^{n_1} &\geq \frac{K_1 ^ {n_1}}{\frac{\beta_1}{\alpha_1} - 1} \nonumber \\
\frac{\beta_1}{\alpha_1} - 1 &\geq \bigg( \frac{K_1}{y(1)} \bigg) ^ {n_1}\nonumber \\
\frac{\beta_1}{\alpha_1} &\geq \bigg( \frac{K_1}{y(1)} \bigg) ^ {n_1} + 1\nonumber \\
\frac{\alpha_1}{\beta_1} &\leq \frac{[y(1)]^{n_1}}{K_1^ {n_1} + [y(1)]^{n_1} } \nonumber \\
- \alpha_1&\geq -\beta_1 \frac{[y(1)]^{n_1}}{K_1^ {n_1} + [y(1)]^{n_1} } \nonumber \\
\alpha_1 - \alpha_1 &\geq \alpha_1 -\beta_1 \frac{[y(1)]^{n_1}}{K_1^ {n_1} + [y(1)]^{n_1} } \nonumber \\
0 &\geq \alpha_1 -\beta_1 \frac{[y(1)]^{n_1}}{K_1^ {n_1} + [y(1)]^{n_1} } \nonumber
\end{align}

We can thus conclude that the mean value theorem condition is satisfied, and there is at least one equilibrium point for the system.

\subsubsection*{A3}\label{A3}
Following the discussion on the transition condition from Region 2 to Region 3 in Section \ref{subsec: aux}, first obtain the trajectory $y(x)$ as follows:

\begin{align}
\frac{dy}{dt} &= \frac{dy}{dx} \frac{dx}{dt} \tag{From the chain rule} \\
\frac{dy}{dx} &= \frac{dy}{dt} \bigg( \frac{dx}{dt} \bigg) ^{-1} \nonumber \\
  &= \frac{ \alpha_2 - \alpha_2 y - \beta_2 y}{\alpha_1 + \alpha_3 -\alpha_3 x} \nonumber \\
\int \frac{1}{ \alpha_2 - \alpha_2 y - \beta_2 y} dy  &= \int \frac{1}{\alpha_1 + \alpha_3 -\alpha_3 x} dx \tag{Separation of Variables}
\end{align}
\[ y(x) = \frac{1}{\alpha_2 + \beta_2} \bigg( \alpha_2 - C( \alpha_1 + \alpha_3 -\alpha_3 x )^{\frac{{\alpha_2+\beta_2}}{\alpha_3}} \bigg)
 \]

where $C$ is a general constant. For any initial conditions $x=x_0, y=y_0$, determine the value of $C$ to be:
\begin{align}
C &= (\alpha_2 - (\alpha_2 + \beta_2)y_0 ) ( \alpha_1 + \alpha_3 -\alpha_3 x_0)^{-\frac{{\alpha_2+\beta_2}}{\alpha_3}} \nonumber
\end{align}

Using the initial conditions $x=K_2, y=0$  to find the corresponding trajectory, obtain
 \[ C = \alpha_2 ( \alpha_1 + \alpha_3 -\alpha_3 K_2)^{-\frac{{\alpha_2+\beta_2}}{\alpha_3}} \]
resulting in
\[ y(x) = \frac{\alpha_2}{\alpha_2 + \beta_2} \bigg( 1 - \bigg( \frac{\alpha_1 + \alpha_3 -\alpha_3 x} { \alpha_1 + \alpha_3 -\alpha_3 K_2} \bigg)^{\frac{{\alpha_2+\beta_2}}{\alpha_3}} \bigg) \]

For the solution to remain in the feasible set and cross the correct border, it must be true that when $y(\hat x)=K_1$, $K_2 \leq \hat x \leq 1$. Therefore, consider the value $\hat x$:

\begin{align}
\hat x &= 1 + \frac{\alpha_1}{\alpha_3} - \bigg( \frac{\alpha_1}{\alpha_3}  + 1- K_2 \bigg) \bigg (\frac{-\beta_2}{\alpha_2} K_1 + 1 - K_1 \bigg)^{\frac{\alpha_3}{\alpha_2+\beta_2}} \nonumber
\end{align}

The transition condition is such that for this value of $\hat x$, $K_2 \leq \hat x \leq 1$.

\subsubsection*{A4}\label{A2}
Following the discussion for the domain in Section \ref{subsec: aux}, obtain $y(x)$ with the initial conditions $x_0=1, y_0=K_1$. From Appendix \nameref{A3},

\begin{align}
C &= (\alpha_2 - (\alpha_2 + \beta_2)K_1)\alpha_1^{-\frac{{\alpha_2+\beta_2}}{\alpha_3}} \nonumber
\end{align}

resulting in the trajectory
\begin{align}
y(x) &= \frac{K_1(\alpha_2-\beta_2)}{\alpha_2 + \beta_2} \bigg( 1 + \frac{\alpha_3}{\alpha_1} (1 - x) \bigg)^{\frac{{\alpha_2+\beta_2}}{\alpha_3}}. \nonumber
\end{align}

Find $\hat x$ such that $y(\hat x)=0$:

\begin{align}
\hat x &= 1 - \frac{\alpha_1}{\alpha_3} \bigg( \frac{\alpha_2}{\alpha_2  - K_1 (\alpha_2 + \beta_2)} \bigg) ^{\frac{\alpha_3}{\alpha_2+\beta_2}} + \frac{\alpha_1}{\alpha_3}  \nonumber
\end{align}

Therefore the condition on the parameters for the domain to be convex and closed is to have \[ K_2 \leq 1 - \frac{\alpha_1}{\alpha_3} \bigg( \frac{\alpha_2}{\alpha_2  - K_1 (\alpha_2 + \beta_2)} \bigg) ^{\frac{\alpha_3}{\alpha_2+\beta_2}} + \frac{\alpha_1}{\alpha_3} \leq 1 \]

or, rewritten:
\[ 0 \leq \frac{\alpha_1}{\alpha_3} \bigg(  \bigg( \frac{\alpha_2}{\alpha_2  - K_1 (\alpha_2 + \beta_2)} \bigg) ^{\frac{\alpha_3}{\alpha_2+\beta_2}} -1 \bigg) \leq 1 \]

\subsubsection*{A5}\label{A7}
For an exponential population model,
\[ \frac{dF}{dt} = rF, \]
where $r$ represents the growth rate of the population, the solution is given by
\[ F(t) = F_0e^{rt}. \]

The doubling time of the population $\tau_2$ is the time it takes for the population to double in size. Therefore,

\begin{align}
F(t + \tau_2) &= 2F(t) \nonumber \\
F(0)e^{r(t+\tau_2)} &= 2F(0)e^{rt}  \nonumber \\
e^{r\tau_2} &= 2 \nonumber \\
r \tau_2 &= \ln(2) \nonumber \\
r &= \frac{\ln(2)}{\tau_2} \nonumber
\end{align}

\end{document}